\documentclass[sigconf,screen]{acmart}

\usepackage{subcaption}
\usepackage{float}
\usepackage{graphicx}
\usepackage{wrapfig}

\copyrightyear{2023}
\acmYear{2023}
\setcopyright{rightsretained}
\acmConference[IUI '23]{28th International Conference on Intelligent User Interfaces}{March 27--31, 2023}{Sydney, NSW, Australia}
\acmBooktitle{28th International Conference on Intelligent User Interfaces (IUI '23), March 27--31, 2023, Sydney, NSW, Australia}
\acmDOI{10.1145/3581641.3584040}
\acmISBN{979-8-4007-0106-1/23/03}

\begin{document}

\title[The Impact of Expertise in the Loop for Exploring Machine Rationality]{The Impact of Expertise in the Loop\\for Exploring Machine Rationality}

\author{Changkun Ou}
\orcid{0000-0002-4595-7485}
\affiliation{
  \institution{LMU Munich}
  \country{Germany}
}
\email{research@changkun.de}

\author{Sven Mayer}
\orcid{0000-0001-5462-8782}
\affiliation{
  \institution{LMU Munich}
  \country{Germany}
}
\email{info@sven-mayer.com}

\author{Andreas Butz}
\orcid{0000-0002-9007-9888}
\affiliation{
  \institution{LMU Munich}
  \country{Germany}
}
\email{butz@ifi.lmu.de}

\renewcommand{\shortauthors}{Ou et al.}

\begin{teaserfigure}
\centering
\includegraphics[width=\textwidth]{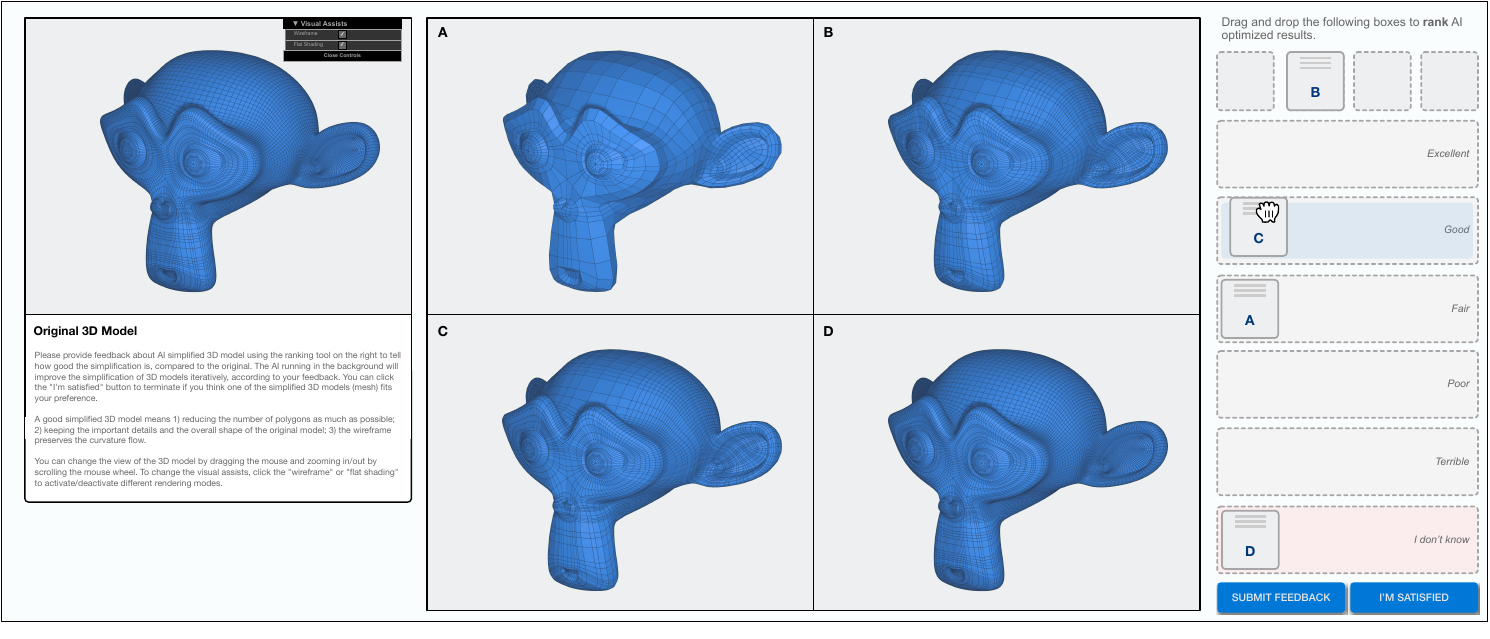}
\caption{The interface for optimizing 3D model simplification using an expert in the loop. In each iteration, the interface presents four 3D models. Participants can drag and drop the top right blocks to a suitable rating region to provide a ranking at submission. Each of the regions can contain multiple blocks. Blocks can be put into ``I don't know'' to express an incomplete preference. Participants can indicate their satisfaction by terminating the optimization loop. To inspect the 3D model quality, they can zoom in/out, pan, and rotate the 3D models simultaneously using a mouse.}
\label{fig:interface-mesh}
\end{teaserfigure}

\begin{abstract}
Human-in-the-loop optimization utilizes human expertise to guide machine optimizers iteratively and search for an optimal solution in a solution space. While prior empirical studies mainly investigated novices, we analyzed the impact of the levels of expertise on the outcome quality and corresponding subjective satisfaction. We conducted a study (N=60) in text, photo, and 3D mesh optimization contexts. We found that novices can achieve an expert level of quality performance, but participants with higher expertise led to more optimization iteration with more explicit preference while keeping satisfaction low. In contrast, novices were more easily satisfied and terminated faster. Therefore, we identified that experts seek more diverse outcomes while the machine reaches optimal results, and the observed behavior can be used as a performance indicator for human-in-the-loop system designers to improve underlying models. We inform future research to be cautious about the impact of user expertise when designing human-in-the-loop systems.
\end{abstract}

\ccsdesc[500]{Human-centered computing~Empirical studies in HCI}
\ccsdesc[300]{Human-centered computing~Interaction paradigms}
\ccsdesc[300]{Computing methodologies~Active learning settings}

\keywords{human-in-the-loop machine learning; adaptive human-computer interaction; rationality}
\maketitle

\section{Introduction}
Human-in-the-loop (HITL)~\cite{monarch2021hitl} optimization uses human expertise to improve machine capabilities. It optimizes system parameters according to human evaluation feedback and supports humans to obtain better outcomes in a variety of areas, such as co-creation~\cite{miller2019creative}, personalized recommendation~\cite{jannach2010recommender}, and decision-making~\cite{bansal2019dms}. When designing a system that involves human feedback, a frequent design decision is not to query absolute human ratings but to ask about the preferred option among a set of design alternatives~\cite{koyam2020seqgallery, koyama2021ui}. If this is done repeatedly, the system iteratively models the user's preferences from the given feedback to infer the next optimal set of options. From a machine perspective, these options represent curiosity regarding what humans might prefer.

When humans face a design problem that requires adjusting various system parameters for the desired outcome, it is tedious to tweak them without prior expertise with the system's behavior. To improve the feedback loop efficiency, one can substitute this process of choosing different parameters with adaptive \emph{exploration} and \emph{exploitation} using human choices. Among many existing approaches, the Bayesian optimization (BO) technique is frequently used and preferred~\cite{brochu2007activelearn, brochu2010animation_design, shahriari2016boreview}. With BO applied, the user interface (UI) can, for instance, present multiple design alternatives, from which users then make a decision~\cite{koyama2021ui}. The system automatically predicts the next best estimations and presents them again based on past choices. The process not only removes the user workload of tweaking parameters but is also expected to propose desired outcomes eventually~\cite{marks1997galleries}. An underlying assumption is that the system outcome could improve when more human expertise is involved in this interaction loop.

However, the system might not always be effective in achieving user satisfaction. There are several known reasons for this, such as context~\cite{rooderkerk2011incorporating}, timing~\cite{holliday2016time}, trustworthiness~\cite{kizilcec2016trust}, cognitive biases~\cite{brochu2007activelearn}, and unstable and contradicting preferences~\cite{ou2022infloop}. Specifically, in co-creation, a user is not always satisfied with the results generated by the machine due to the lack of practical creativity in the system~\cite{miller2019creative}; in a personalized recommendation, the machine may converge to some fixed recommended content and cannot bring fresh ideas for users~\cite{jannach2010recommender}; even in the process of AI-assisted decision-making, users may not hold enough trust in the results provided by an algorithm~\cite{dietvorst2015algaversion}.

Although there are strategies to mitigate these subjective imperfections on the human side, such as improving transparency~\cite{kizilcec2016trust}, interpretability~\cite{ehsan2021explain}, and control~\cite{yijun2021melody}, the reported dissatisfaction, lack of freshness, and trust remain subjective and are measured exclusively using subjective scales, Moreover, empirical studies also mainly report based on novice user groups~\cite{koyam2020seqgallery, chan2022novice}. The impact of the involved expertise on the overall system outcome quality is rarely discussed. On the other hand, we can not easily assess the objective outcome quality reliably if the results partially depend on subjective concepts. Since the expertise involved plays an important role in the obtained human feedback, we investigate the following two research questions:

\textbf{RQ1}~\emph{How do HITL optimization outcomes differ objectively when using preferential feedback from humans?}, and

\textbf{RQ2}~\emph{What is the impact of the involved user expertise on the system outcomes and subjective satisfaction?}

\noindent
In particular, we are interested to see \emph{how the answers to these questions could provide insights for designing future HITL systems}. To cover a spectrum of different application domains, we consider text summarization~\cite{simpson2020interactive}, photo color enhancement~\cite{koyama2016selph, koyama2017sequential, koyam2020seqgallery}, and a 3D model simplification task~\cite{garland1998quadrics, jakob2015instant, ou2022infloop} to evaluate the relation between user expertise, satisfaction, and system outcome quality when interacting with an intelligent system. \autoref{fig:interface-mesh} shows one of our ranking interfaces for HITL optimization. Our selected tasks are not only challenging to design algorithms for and to solve technically, as they require not only objective measurements but also subjective opinions~\cite{arvo2015survey, lescoat2020specmeshsim}. Therefore, we conducted a study to collect choice behaviors in a user group (N=60) with different levels of expertise in three contexts to assess the overall interaction and optimization loop. We also asked about their subjective satisfaction regarding the final system outcomes.

As a key result, we evaluate the connection between user expertise, subjective satisfaction, and the quality of the system outcomes in an interactive feedback loop. Our evaluation indicates that novice subjects can produce an equal outcome quality even faster (and be more satisfied with it) than those with higher expertise. The main contribution of this paper is an empirical investigation of the impact of involved human expertise on the overall HITL optimization performance. We also discuss design implications and potential future directions to consider the impact of user expertise in  HITL optimization applications.

\section{Background and Related Work}
To start, we briefly discuss the state-of-the-art approaches for modeling human feedback iteratively using Bayesian optimization. Then, we overview prior literature in the social sciences, mainly psychology, to resolve the potential ambiguity regarding the terms ``satisficing'' and ``expertise'' and approaches to quantify them. They serve as the foundation of our problem description and support our assessment of user expertise and satisfaction in HITL settings.

\subsection{Modeling Preference from Human Feedback}

\begin{figure}[b]
\centering
\includegraphics[width=\columnwidth]{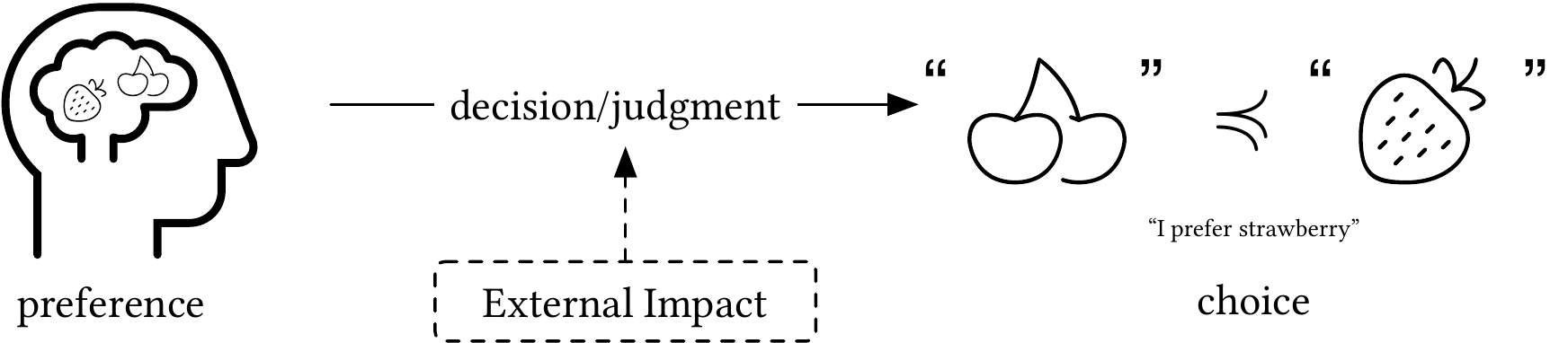}
\caption{A decision process turns an internal preference state into an observable choice. The choice may not reflect the underlying preference due to external influences.}
\label{fig:prefdef}
\end{figure}

Human-in-the-loop optimization outcomes depend on the machine algorithm capability as well as the preferential feedback expressed by a human user. Studies on the term ``preference'' appear in many disciplines. For clarity in the subsequent discussions, in this paper, we follow~\citet{hausman2011preference} in their counterargument against eliminating preference using choice~\cite{gruene2004revealed} and acknowledge the existence of \emph{preference}. In our use of terms, shown in~\autoref{fig:prefdef}, preference is a subjective concept representing an impermeable and unobservable state of an individual mind. A preference may or may not be present when the individual encounters multiple given \emph{options}. A \emph{choice} denotes the objectively observable actions of the individual that selects at least one option among the given ones, and \emph{decision} or \emph{judgment} reveals a subjective realization process from a preference towards a choice. A choice may not reflect the underlying preference due to external influences.

In existing theories regarding preferences in psychology and economics, theoretical models tend to infer preference from comparisons~\cite{thurstone1927comparelaw} and rely on basic axioms~\cite{anand1987axioms} of this preference logic: \emph{completeness} and \emph{transitivity}. The completeness axiom assumes the existence of preference, which guarantees that individuals can always express their preference by choosing among at least two options; transitivity means that we can infer that A is preferred over C if we prefer A over B and B over C.

Although these axioms are convenient for a rigorous discussion of the logic of preferences, they are still strong assumptions that may easily be violated. Behavioral literature shows that choices are partly dominated by the context~\cite{rooderkerk2011incorporating}, and the transitivity axiom is not applicable when implicitly involving other judgments that were not previously considered. For example, when a human prefers A over B and B over C involving only one objective, they may implicitly involve another, previously unconsidered objective when choosing between A and C as a pair. As a result, C may be preferred over A. Moreover, the completeness axiom may be violated when the human thinks that ``I don't know'' or ``I don't care'' among two subjectively indistinguishable options, thus causing a \emph{random choice}.

Since BO learns a \emph{posterior} from human feedback, it aims to search for a maximum of an unknown function by exploring and exploiting the solution space. Therefore, it can propose examples using an \emph{acquisition function}, ask the human to provide a choice, and then infer the underlying preference iteratively.
When dealing with choices from pairwise comparisons,
preferential Bayesian optimization (PBO)\footnote{We use PBO as a more general term to represent a category of methods that infer preference from choice, in contrast to the specific approach by~\citet{gonzalez2017pbo}.}
as a specialized category of BO has received increasing development in recent studies~\cite{gonzalez2017pbo, koyam2020seqgallery, mikkola2020projective, siivola2021pbbo, lin2022preference}. While BO learns based on absolute rating utility (rate and assign a score to an option), PBO learns from human choice in pairwise comparisons according to Thurstone's law of comparative judgment~\cite{thurstone1927comparelaw}. To avoid the mentioned violations of the transitivity axiom, the recent extensions~\cite{koyam2020seqgallery, siivola2021pbbo, benavoli2021choice} to PBO transited from using a binary pairwise comparison to using a reasonable amount of options. These extensions can largely prevent violation of the transitivity axiom and infer more information at a time because they either consider choosing a set of options as winners among all given options~\cite{koyam2020seqgallery, benavoli2021choice}; or provide a ranking of all given options, where options may share the same level of rank~\cite{siivola2021pbbo}. Note that more ranking elements may also increase the uncertainty for users to make imperfect decisions~\cite{monarch2021hitl} due to increased workload. Thus, one should carefully consider the presented number of elements.

Although PBO has used pairwise comparisons to mitigate various issues regarding unstable human judgments, these current approaches could also violate preference axioms due to cognitive limitations. \citet{kahneman1974judgment} have widely presented how heuristic biases might influence the choice behavior. External causes can also produce a considerable amount of noise in choice~\cite{kahneman2021noise}.
To overcome the violation of preference axioms, PBO has also considered handling noisy inputs~\cite{letham2019constrained} and guarantees theoretical convergence when dealing with unstable choices. Despite all these developments in PBO, we still observe two challenges in practice: The first challenge is that a human might change their objective during the integrative optimization, even using pairwise comparisons, because PBO assumes a fixed implicit underlying choice function which it can learn. Although PBO can deal with noisy inputs, another challenge is that it requires much more iterations to let the optimization converge. This is usually very costly when involving a human, and we also need to design the UI carefully to mitigate these issues and reduce user errors.

Therefore, there are several major design considerations to obtain reliable preference from choice: 1) the underlying learning algorithm should effectively deal with feedback uncertainty and noisy input of an individual being, 2) the objective of the user task should be provided to avoid incomplete preference, and 3) The user interfaces should support users to express their incomplete preferences explicitly.

\subsection{Bounded Rationality and Satisficing}

Understanding the satisfaction of users when they are involved in a loop requires deeper insights from human psychological factors regarding bounded rationality and satisficing.
\citet{simon1955rational} first coined the term \emph{bounded rationality} to describe the perceived information limits individual rationality. This observation provides a sufficient discussion base for interpretations regarding irrational decisions. As previously discussed, \citet{kahneman1974judgment} emphasized one possible category of systematic errors from the cognitive perspective. In recent discussions~\cite{brandstaetter2006tradeoff,gigerenzer2009homo}, researchers take a statistical perspective and underline that recurring noise could also contribute equally~\cite{kahneman2021noise} to bounded rationality.
This is met by matching behavior from the preference point of view, as bounded rationality appears in the decision or judgment process and causes the violation of the completeness axiom due to \emph{satisficing}.

Satisficing is a ``good enough'' decision strategy~\cite{simon1955rational, schmidtz2004satisficing} that ends the search process when a certain threshold quality is met. When some of the presented options are subjectively acceptable, the effects of bounded rationality and satisficing cause the process to terminate with a sub-optimal outcome. An opposite decision tendency is called \emph{maximizing}, where a final decision cannot be made without enough information.
\citet{schwartz2002maximizing} provided evidence by assessing subjective happiness and individual differences in what people aspire to when they make decisions in various domains of their lives. People who use a maximizing strategy desire the best possible result. Although the authors did not find any strict causality for a maximizing strategy producing significantly lower satisfaction with life than satisficing, they argue that a maximizing decision strategy might constantly look for better objective outcomes.
In modern recommender systems, for example, prior work~\cite{iyengar2006doing} showed that a satisficing strategy leads to quicker selections and increased content viewing time. In contrast, subjects using a maximizing strategy spent significantly more time on selection activities. In comparison, subjects using satisficing decision strategies spent significantly less viewing time, regardless of subjective content quality.

The objective reasons why HITL optimization systems work differently for bounded rational human agents remain underexplored. Although previous psychology research correlated rationality with using satisficing and maximizing decision strategies, there is little discussion about what objective properties lead to the reported subjective dissatisfaction in this new context. Especially as the previously reported unsatisfactory results rarely evaluate the objective quality of optimized choice options while involving different levels of rationality, we also wonder if an unsatisfactory result has comparably lower quality or whether satisficing is sufficient to maximize a machine learner's capability. In addition, with proper selection on a task, the quality of a rational choice is also part of the consequence of human intelligence, known as ``expertise''. Concerning decisions using expertise, empirical research also reports that people with high expertise apply more criteria during their decision, especially clinic decisions~\cite{grove1996comparative}, which proved less efficient and more correlated to a maximizing strategy. Still, it remains unclear what the satisfaction would be in this case.

\subsection{Expertise in Context}
\label{sec:expertise}
To analyze the concept of expertise and quantify the impact of involved expertise, one of the most straightforward questions regarding expertise is: ``what is an expert?'' Depending on the domain context, there are different decompositions of the concept of expertise. In particular, \citet{garrett2009dim6} describe six different dimensions regarding expertise, whereas \citet{collins2013dim3} suggests three dimensions and \citet{kotzee2018dim2} suggest only two dimensions based on social aspects. On a higher level, \citet{bourne2014def} argues for interpreting expertise as a descriptive term that involves knowledge and skills, which are mental or cognitive concepts rather than physical talent. Therefore, tasks that might be physically quickly adapted and measured regarding efficiency are less suited to verifying the expertise involved.

To quantify the loosely defined concept ``expertise,'' a range of theoretical models have been developed, e.g., by describing a game between a decision-maker and an expert who proposes options~\cite{krishna2001model}.
For our purpose here, we are interested in quantifying the level of expertise of a specific human within a particular context.
\citet{treem2016communicate} propose to define 1) an observer who knows what it looks like and 2) an expert who has an objective communicative skill that outperforms the observer who can infer their expertise. \citet{ooge2021trust} further developed this concept and introduced a third metric for inferring expertise by using a preliminary task to measure a person's performance.

Because of the interpretation ambiguities and different arguments about proper measures in other contexts, instead of asking about an absolute level ``is A an expert?'', identifying a person with a comparatively higher level of expertise than another appears to be a more reliable local assessment. This transition turns the expertise assessment into a  ranking question ``is A better than B in context C?'' similar to preference ordering~\cite{lee2012ranking}.
\citet{ferrod2021identifying} turned the problem of detecting the level of expertise of a user from dialogues into a text classification task that concerns and emphasizes expertise in the telecommunication domain. Although their measures are not directly transferable to a general context, the classification methodology confirms that \emph{relative} expertise inferred from classification can avoid defining absolute levels. The literature analysis in this section identified that expertise is highly context-dependent, and that human experience is relevant. To measure the involved expertise in a feedback loop, one does not only need to measure the human experience but also needs to consider the context involved.

\section{User Study}

We designed the following user study to answer our research questions (\textbf{RQ1} and \textbf{RQ2}).
To understand the impact of expertise on satisfaction, we hypothesize that by using HITL optimization, participants with a higher level of expertise will produce a better outcome quality and perceive higher satisfaction than novice participants. To verify this, we designed a between-group controlled experiment in three problem contexts: text summarization, photo color enhancement, and 3D model simplification. As dependent variables, we measured participants' expertise in a domain context, interactions with the system, and feedback from final questionnaires (individual rating scales and open questions).

\begin{figure*}[t]
\begin{subfigure}[b]{\linewidth}
         \centering
         \includegraphics[width=\linewidth]{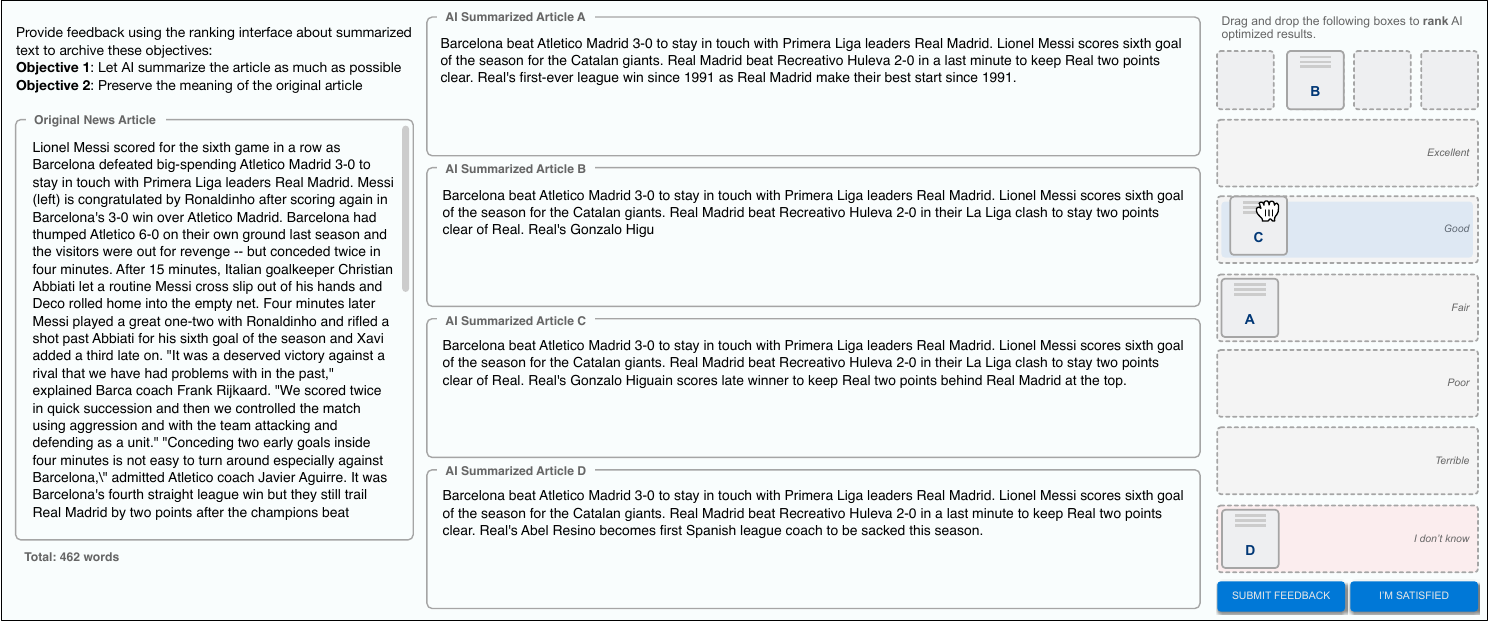}
         \caption{AI-based text summarization.}
         \label{fig:interface-text}
     \end{subfigure}
     \begin{subfigure}[b]{\linewidth}
         \centering
         \includegraphics[width=\linewidth]{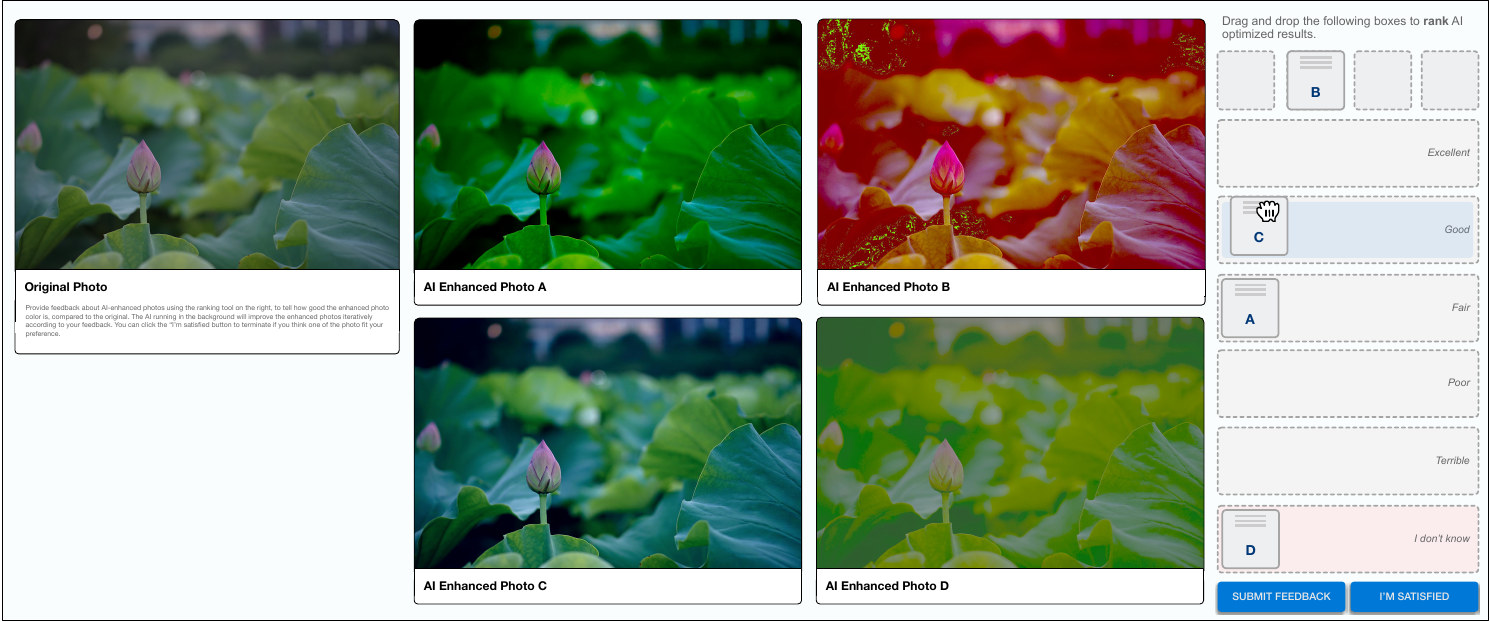}
         \caption{AI-based photo color enhancement. The photo is taken from~\citet{koyam2020seqgallery}.}
         \label{fig:interface-image}
     \end{subfigure}
\caption{The ranking interface for a) text summarization and b) photo color enhancement. In each iteration, the interface presents four options. Participants can drag and drop the top right blocks to a suitable rating region to provide a ranking of the options regarding the given objectives. Each of the regions can contain multiple blocks. Blocks can be put in the ``I don't know'' region to express an incomplete preference.}
\label{fig:interfaces}
\end{figure*}

\subsection{Problem Context}
In a HITL optimization context, it is more fitting to use decision-making tasks that sit between pure subjective preference matter (e.g., favorite colors) and well-defined objective optimization problems that can be solved procedurally (e.g., finding the global minimum of a strict convex continuous function). We need to select tasks where users provide ranking feedback using their expertise. A task should also be iterative for observing progress and partially subjective because users could balance the trade-off on different objectives.

We selected tasks that include text summarization, photo color enhancement, and 3D model simplification for the following reasons:
1) They all partially involve rational, objective judgment, and subjective components.
2) Each domain requires different levels of human expertise: text summarization only requires language proficiency, which is a fundamental human expertise; photo color enhancement requires an understanding of aesthetics and color theory; 3D model simplification requires domain-specific technical 3D modeling expertise.
3) All these contexts had been discussed in the HITL optimization literature~\cite{simpson2020interactive, koyama2016selph, koyama2017sequential, koyam2020seqgallery, ou2022infloop} individually but not compared to each other together.

\subsection{User Interfaces for Data Collection}

\autoref{fig:interface-mesh}, and \ref{fig:interfaces} show our UIs in the HITL optimization main task for 3D model simplification, text summarization, and photo color enhancement, respectively.
All interfaces collect a participant's expertise at the beginning of the study, then present four variants through the interface. When a task is over, the interface presents six final questions and an open question regarding their satisfaction and overall experience when interacting with the system.
In all system interfaces, users can express their ranking choices, and users provide a ranking of the current four result variants on the interface's right side. Additionally, in the 3D model simplification task, a user can rotate, zoom, and move the four models simultaneously to inspect and compare the quality of the models.
In line with prior work by \citet{ou2022infloop}, we use a listwise interface with four variants instead of two pairwise comparisons to increase the collected feedback in each iteration without increasing system processing and data transmission time. After the user submits the ranking information, the background system will utilize this information and then optimize and infer the next optimal set of variants.
We also added an ``I don't know'' container box to the ranking UI and allowed participants to express incomplete preferences. This design is intended to prevent the violation of the completeness axiom.

\subsection{Apparatus}
We developed the frontend UIs using Material UI\footnote{\url{https://mui.com/}, \emph{last accessed \today}}, React DnD\footnote{\url{https://react-dnd.github.io/react-dnd/about}, \emph{last accessed \today}}, and three.js\footnote{\url{https://threejs.org}, \emph{last accessed \today}}.
Apart from the frontend, our backend \emph{core service} is written in Go\footnote{\url{https://go.dev}, \emph{last accessed \today}} for easier concurrency management. It serves our frontend interfaces, data collection, and communications with other dedicated computing microservices, including \emph{domain services} and an \emph{optimizer service}. The logged data were directly managed using the OS file system with naming conventions. We deployed all services on our institute infrastructure (Ubuntu 20.04, 8-Core Intel Core i9-9900K, 64GB RAM, and NVIDIA GeForce RTX 2080 Ti with 11GB of GPU memory).

\begin{figure*}[t!]
\centering
\includegraphics[width=\textwidth]{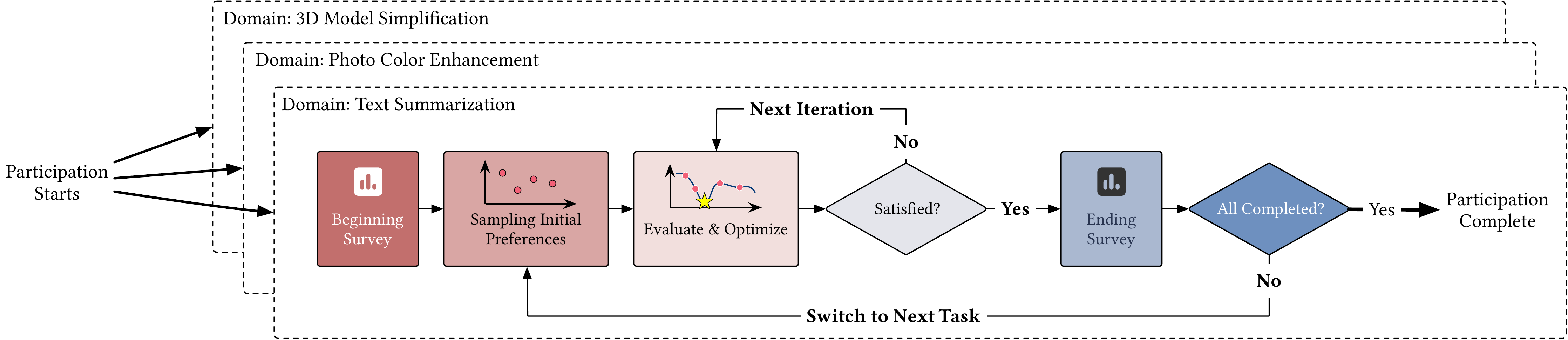}
\caption{The study procedure: Each participant did one of the three domain contexts (news articles, photos, and 3D models).}
\label{fig:procedure}
\end{figure*}

\subsubsection{Domain services}
We implemented three separate domain services. To perform text summarization, we picked the pre-trained BART model via HuggingFace\footnote{\url{https://huggingface.co/sshleifer/distilbart-cnn-12-6}, \emph{last accessed \today}}. We implemented an isolated text summarization server using Flask\footnote{\url{https://flask.palletsprojects.com/en/2.2.x/}, \emph{last accessed \today}} with GPU acceleration.
We use Nucleus sampling~\cite{holtzman2020topp} as a stochastic text decoding strategy\footnote{\url{https://huggingface.co/blog/how-to-generate}, \emph{last accessed \today}\label{footnote:generate}} for our inference use case because it allows for a bounded hyper-parameter space (between 0 and 1) and can generate diverse human-like sentences in the inference phase. Since we designed our user task to consider the length of summarization as a decision criterion, we used a summarization ratio as a hard limit that controls the text generation length and a length penalty as a selected soft limit that encourages the model to generate shorter text. As a result, our hosted text summarization service allows four adjustable system hyper-parameters at every model inference stage:
1)~\emph{summarization ratio},
2)~\emph{length penalty},
3)~\emph{top-p}, and
4)~\emph{temperature}.

For photo color enhancement, we used a parameterized photo enhancer~\cite{koyama2016selph, koyama2017sequential, koyam2020seqgallery} as an image processing service for better integration to the core service. This service allows five adjustable system hyper-parameters that are bounded between 0 and 1:
1)~\emph{brightness},
2)~\emph{contrast},
3)~\emph{saturation},
4)~\emph{temperature}, and
5)~\emph{tint}.
Lastly, we used a 3D mesh reducer~\cite{ou2022infloop} as a 3D mesh processing service, and it also contains five bounded system hyper-parameters:
1)~\emph{simplification ratio},
2)~\emph{border preservation},
3)~\emph{hard edge preservation},
4)~\emph{sharpness preservation}, and
5)~\emph{quadrilateral preservation}.

\subsubsection{Optimizer service}
\label{sec:bo-service}
We implemented the underlying optimizer using BoTorch~\cite{balandat2020botorch} as a command line service, which reads the user responses to estimate the next optimal system hyper-parameters for exploration. BoTorch provides the EUBO~\cite{lin2022preference} optimizer as one of the state-of-the-art PBO optimizers that consider noisy inputs to estimate system hyper-parameters for pairwise comparisons.
We extended EUBO to utilize ranking comparisons to fit a Gaussian process using the user's rank data first. Then, we used the learned latent utility value to fit another Gaussian process and infer the next batch of exploration positions.

\subsection{Procedure}
\begin{figure*}[t]
\begin{subfigure}[b]{\linewidth}
         \centering
         \includegraphics[width=\textwidth]{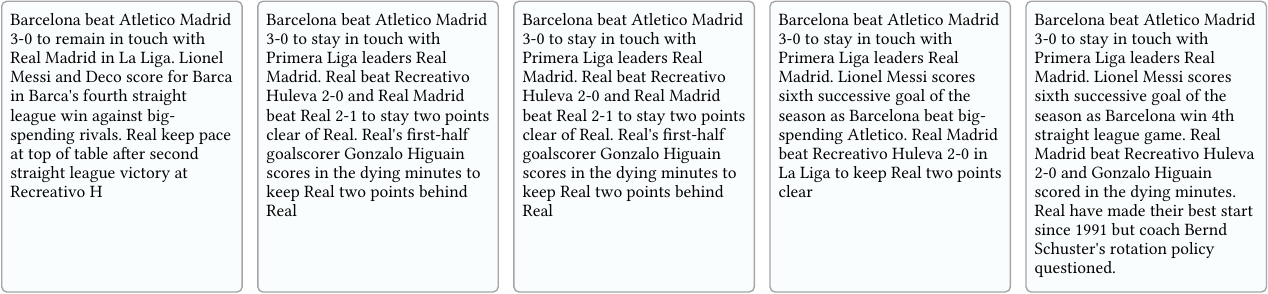}
         \caption{AI-based text summarization.}
         \label{fig:outcomes-text}
     \end{subfigure}
     \begin{subfigure}[b]{\linewidth}
         \centering
         \includegraphics[width=\textwidth]{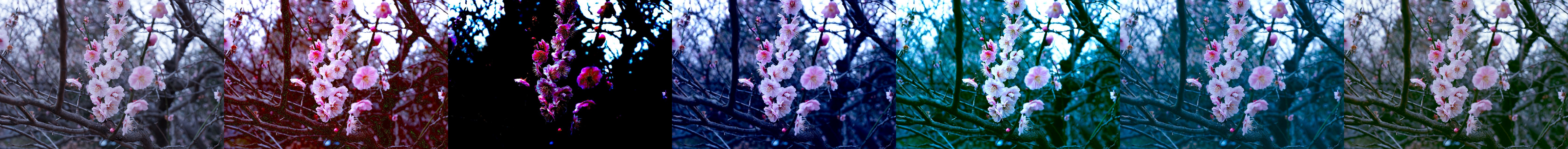}
        \includegraphics[width=\textwidth]{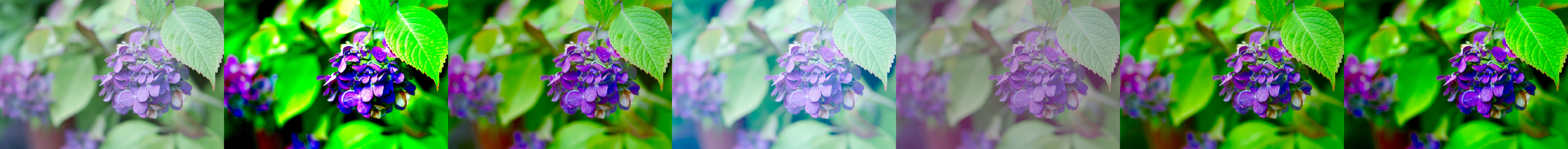}
        \includegraphics[width=\textwidth]{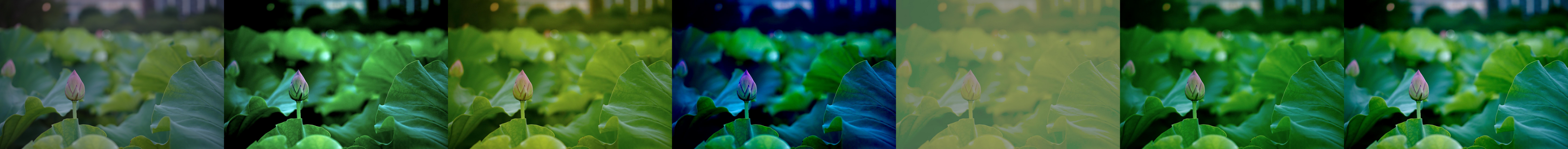}
         \caption{AI-based photo color enhancement. Original photos are taken from~\citet{koyam2020seqgallery}}
         \label{fig:outcomes-image}
     \end{subfigure}
    \begin{subfigure}[b]{\linewidth}
         \centering
        \includegraphics[width=\textwidth]{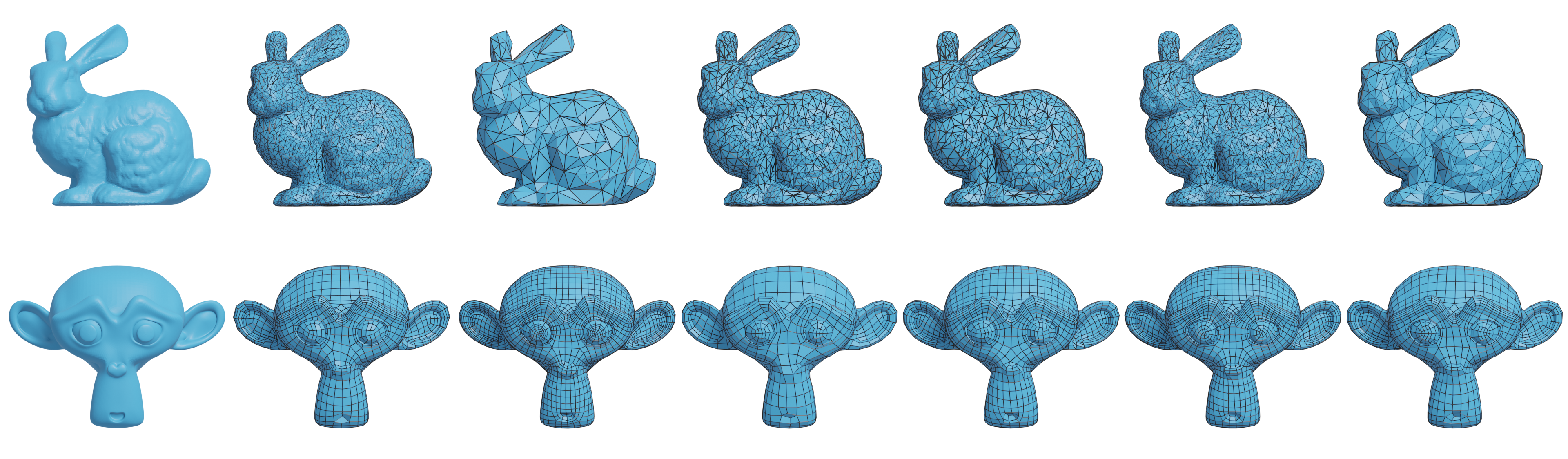}
         \caption{AI-based 3D model simplification.} %
         \label{fig:outcomes-mesh}
     \end{subfigure}
\caption{Example outcome sequences from the a) text summarization,  b) photo color enhancement, and c) 3D model simplification. From left to right, it shows how the objective was optimized progressively until the final satisfying outcome (far right).}
\label{fig:outcomes}
\end{figure*}

The overall procedure is visualized in~\autoref{fig:procedure}. Participants started the study with an informed consent form and answered initial demographic questions, including their age, gender, and domain expertise.
The UI presents a set of evaluation options to participants. The main task is to provide feedback, using the UI, to the AI to 1)~for a news article: \emph{summarize the given news articles as much as possible while preserving the meaning of the article}; 2)~for a photo: \emph{improve and enhance the color of the photo}; and 3)~for a 3D model: \emph{simplify the number of polygons as much as possible while preserving the overall appearance}.
Because the optimizers need initialization samples, in the first 4 iterations, a participant evaluated outcomes produced by quasi-random Sobol sampled~\cite{rezende2014stochastic, wilson2017reparam} system hyper-parameters.
After acquiring these preference priors from participants, starting from the 5th iteration, the optimizer is used, and participants can freely terminate if satisfied with the current text summary. The task automatically ends after 20 iterations to limit participation time.
After termination, participants answered six questions regarding their satisfaction with the system outcomes and their experience using the interface to give feedback.
Each participant completed 3x3 Latin square-shuffled news articles\footnote{Selected from the CNN daily mail dataset, article IDs: ea06fd0b25cb9793397a, 35f0e33de7923036a97a, 42c027e4ff9730fbb3de. See \url{https://huggingface.co/datasets/cnn_dailymail}, \emph{last accessed \today}},
photos\footnote{Selected from~\citet{koyam2020seqgallery}. See \url{https://github.com/yuki-koyama/sequential-gallery/tree/main/resources/scaled}, \emph{last accessed \today}}, and 3D models\footnote{Selected model name: stanford bunny, suzanne, and fandisk. See~\url{https://github.com/alecjacobson/common-3d-test-models}, \emph{last accessed \today}}. Example iteratively optimized outcomes are shown in~\autoref{fig:outcomes}.

\subsection{Participants}
We recruited participants worldwide on Prolific.
Because participants had different median completion times in different experimental conditions, we paid between £3 to £9 upon completion, corresponding to an hourly wage of £9/h (\$10.4/h). Participants gave informed consent at the beginning of the study; thus, the study adhered to European privacy laws (GDPR). In total, we collected data from 91 participants from 13 countries.
To guarantee high-quality responses, we only consider participants if they: 1)~had an approval rate of 95\%, 2)~completed the study only once, 3)~answered with consistent demographics, e.g., not more than five years of age difference in the study compared to the platform registration information, and 4)~provided their response in at least a reasonable amount of time, i.e., spent longer than 3 seconds in each iteration to read the summarized text and interact with the interface according to our pilot study observations.
Therefore, we will report our results based on 60 participants (31 female, 29 male, and no diverse; age $\mu=26.92, \sigma=6.44$, range 19-52). Each domain context includes 20 participants.

\subsection{Measurements}
During the study, we measured participants' expertise, their interaction behavior with our developed system, subjective ranking feedback to the system outcomes, objective quality of system outcome, and their subjective satisfaction and open questions regarding their experience.

\subsubsection{Expertise Measures}
\label{sec:infer}
As discussed in \autoref{sec:expertise}, since user expertise is measured differently in prior research, we use a similar approach as~\citet{ooge2021trust, ferrod2021identifying}, and combine the following established metrics: 1)~self-indication, 2)~accumulated work experience, and 3)~familiarity with domain problems.
For the text context, we ask for their language proficiency, and for the image and mesh contexts, we ask for their self-indicated photo editing and 3D modeling expertise. All contexts asked for their months of work experience as well as when was their last time of experience.
Based on the collected data, relative levels of expertise are used in our context for the discussion of expertise, and we normalized these measures among all participants, then grouped participants into three groups using quantile-based discretization: \emph{novice}, \emph{intermediate}, and \emph{experienced}. Note that the descriptions only represent the relative levels among our recruited participants. In a larger user group, they may be reconsidered as novice or intermediate accordingly.

\subsubsection{Behavior Measures}
We measured how participants interacted with our developed systems, including
1)~decision time: the period between the appearance of the evaluation options and a full ranking is submitted,
2)~number of iterations per task,
3)~number of incomplete preferences per iteration,
4)~number of indifference preferences per task, and
5)~number of drag and drop operations to rank given options.

\subsubsection{Objective Outcome Quality Measures}
\label{sec:metrics}
In the text summarization context, we measured the total number of words in the outcome texts to validate if participants made progress on the given objectives. We also computed the BLEU~\cite{papineni2002bleu} and ROUGE~\cite{lin2004rouge} (including ROUGE-1, ROUGE-2, and ROUGE-L) scores to measure the objective quality of summarized texts as they are frequently used to evaluate the quality of summarization, and correlate positively with human evaluation.
For the photo color enhancement context, we converted the outcome photo from RGB color space to HSV and YUV to directly reflect the relevant optimization metrics. Namely, we computed saturation, contrast, and brightness using a mean of pixel-wise subtraction between source and outcome in the H, S, and V channels correspondingly. Furthermore, we computed the mean pixel-wise difference of U channels in YUV color space for tint changes and similarly in the V channel for temperature changes.

The task of 3D model simplification concerns simplification ratio~\cite{botsch2010polygon} and perceivable changes regarding visual quality, wireframe quality, and surface quality. The visual quality and wireframe quality are useful indicators concerning human perceptual judgments. In contrast, surface quality is defined at a technical level and was found to be more difficult to perceive visually compared to the other qualities~\cite{corsini2013perceptual}.
Therefore, we computed the simplification ratio to validate whether participants progressed on the given objectives.
In terms of visual quality, one can use rendering quality from multiple camera views to measure visual quality during mesh simplification. A 3D model can be rendered as a series of images from different perspectives with given rendering settings, such as specified light conditions, camera settings, and rendering algorithms. We use an equally weighted combination of \emph{peak signal-to-noise ratio}~\cite{korhonen2012peak} (PSNR) and \emph{structural similarity}~\cite{wang2004ssim} (SSIM) to measure the rendering difference.
For wireframe quality, we computed \emph{scaled Jacobian cell quality}~\cite{knupp2000fem} because it was previously suggested to better correlate with the human judgment~\cite{corsini2013perceptual}. The scaled Jacobian cell quantity itself measures how a given face is regularized.
Lastly, we sampled two point clouds on the source mesh and the outcome, then used \emph{Chamfer distance}~\cite{barrow1977chamfer} to indicate the surface quality. Surface quality is less observable compared to the other three objectives.

\subsubsection{Subjective Measures}
\label{sec:satisfy}
After participants ended a task, we asked six questions: Q1)~participants' overall subjective satisfaction with the final results; Q2)~ their confidence if they think they can do a better version by themselves than the system, which was optimized based on their provided feedback; Q3)~whether the final outcome matched their expected result; Q4)~whether they felt improvements of the result from iteration to iteration; Q5)~whether they felt the ``I don't know'' option was useful, and Q6)~whether they believed they gave clear feedback to the AI. We measured these questions using a bipolar slider-based Likert scale. Among these questions, Q1 to Q4 are intended to measure subjective satisfaction.

\section{Results}
\label{sec:results}

Based on the behavior and subjective questionnaires, we first verify the relationship between user expertise and satisfaction. Then, we evaluate the optimization loops in different domains based on the collected ranking data and objective quality measurements of outcomes.

\subsection{Behavior and Subjective Satisfaction}

To analyze the behavior and subjective satisfaction, we first group our participants using quantile-based discretization to guarantee each grouped expertise level has an evenly distributed number of participants. Then we assert the data's normality using the Shapiro-Wilk test. We use a t-test to compare the difference between novices and experienced participants for normally distributed data. Otherwise, we report a Wilcoxon rank sum test as a non-parametric approach to compare the differences between novice and experienced participants for measured dependent variables. Full results are available in~\autoref{sec:open-science}.

\subsubsection{Inferred Expertise}

Our participants reported varied experiences in different domains. They self-indicated English proficiency on the CEFR scale\footnote{CEFR scale: \url{https://www.coe.int/en/web/common-european-framework-reference-languages/table-1-cefr-3.3-common-reference-levels-global-scale}, \emph{last accessed \today}}: B1 10.00\%, B2 30.00\%, C1 35.00\%, and C2 25.00\%.
For self-indicated expertise in photo editing: none 25.00\%, novice 45.00\%, intermediate 25.00\%, experienced 5.00\%, experts 0.00\%.
For self-indicated expertise in 3D modeling: none 35.00\%, novice 45.00\%, intermediate 15.00\%, experienced 5.00\%, and none indicated themselves as experts.

Participants indicated their period of work experience. For text summarization: No work experience 25\%, less than one year of experience 30\%, 1 to 5 years 25\%, more than 5 years 20\%; for photo editing: No work experience 50\%, less than one year of experience 10\%, 1 to 5 years 30\%, more than 5 years 10\%; for 3D modeling: No work experience 60\%, less than one year of experience 40\%.
Regarding the recent experience in these domains, for text summarization: Never 5\%, in recent 2 weeks 20.0\%, 2 weeks to 3 months ago 25.0\%, 3 to 6 months ago 10.0\%, 6 to 12 months ago 20.0\%, 13 to 36 months ago 5.0\%, more than 36 months ago 15.0\%. for photo editing: Never 10.0\%, in recent 2 weeks 40.0\%, 2 weeks to 3 months ago 30.0\%, 3 to 6 months ago 5.0\%, 6 to 12 months ago 10.0\%, 13 to 36 months ago 5.0\%; and for 3D modeling: Never 40.0\%, in recent 2 weeks 15.0\%, 2 weeks to 3 months ago 15.0\%, 3 to 6 months ago 5.0\%, 6 to 12 months ago 5.0\%, 13 to 36 months ago 5.0\%, more than 36 months ago 15.0\%.

In total, using quantile-based discretization, we inferred participants' level of expertise in the three contexts: text summarization (Novice: 7, Intermediate: 7, Experienced: 6); photo color enhancement (Novice: 7, Intermediate: 7, Experienced: 6); 3D model simplification (Novice: 7, Intermediate: 6, Experienced: 7).

\subsubsection{Interaction Behaviors}
\begin{figure*}[t]
    \centering
    \includegraphics[width=\textwidth]{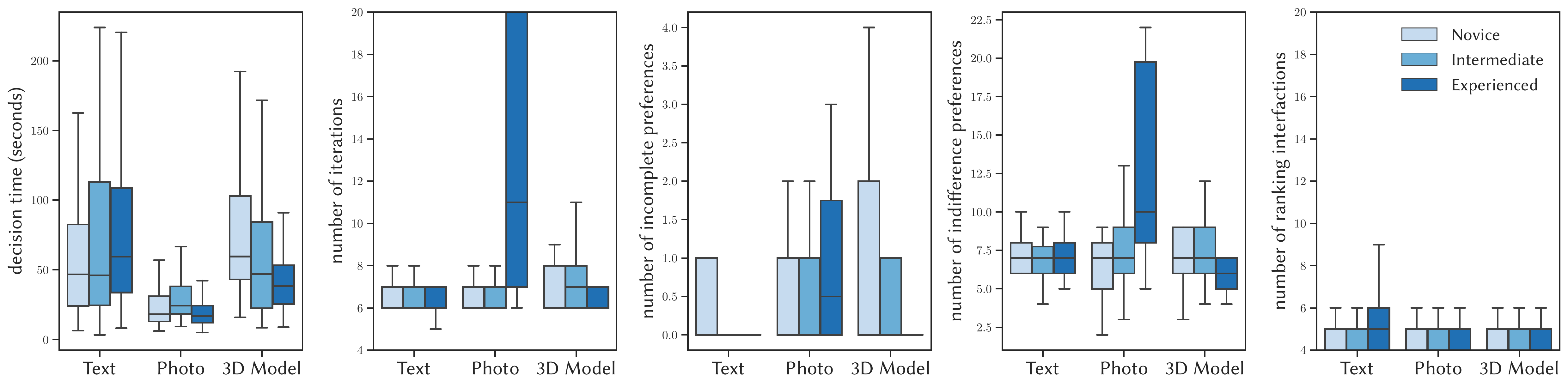}
    \caption{Measured interactions of participants in different domain contexts. Measurements are grouped by the level of expertise. The results indicate that experienced participants express their preferential ranking decisions more clearly than novices. For example, they behave faster in decision time with more iterations or decide slower with more ranking interactions (thoughtful indecision); they also express fewer incomplete and more indifferent preferences.}
    \label{fig:behaviors}
\end{figure*}

All measured interaction behavior indicators are visualized in ~\autoref{fig:behaviors}. In terms of the decision time, we found a significant difference between novices and experienced participants both in text summarization (W = 8273.00, p = .051; r = -0.14, $CI_{95\%}$=[-0.27, -0.0006]), photo color enhancement (W = 22320.00, p = .041; r = 0.12, $CI_{95\%}$=[0.006, 0.23]), and 3D model simplification (W = 20999.50, p < .001; r = 0.45,  $CI_{95\%}$=[0.34, 0.54]). This means \emph{experienced participants are either more thoughtful (e.g., in the text summarization domain) or more effective (e.g., photo color enhancement and 3D simplification) in forming their decision}.
For the number of involved iterations, we did not find a significant difference between novices and experienced participants in text summarization (W = 223.00, p = .953; r = 0.01, $CI_{95\%}$=[-0.33, 0.35]) and 3D model simplification (W = 199.50, p = .751; r = 0.06, $CI_{95\%}$=[-0.30, 0.40]). However, we found significant more iteration in photo color enhancement (W = 99.00, p = .008; r = -0.48, $CI_{95\%}$=[-0.71, -0.15]) for experienced participants than novices. The results suggest that \emph{experienced participants explore the solution space significantly more when the feedback loop is more efficient}.

When checking the expressed number of incomplete preferences, we found experienced participants rarely express an incomplete preference, and novices in the 3D model simplification domain express incomplete preference significantly more than experienced participants (W = 249.00, p = .023; r = 0.32, $CI_{95\%}$=[-0.04, 0.60]) domains. However, we did not find a significant difference in text summarization (W = 274.50, p = .081; r = 0.24, $CI_{95\%}$=[-0.10, 0.54]) and in photo color enhancement (W = 144.50, p = .154; r = -0.24, $CI_{95\%}$=[-0.54, 0.13]) contexts.
Similarly, we found experienced participants indicated indifference preference significantly more than novices in the photo color enhancement domain (W = 102.50, p = .015; r = -0.46, $CI_{95\%}$=[-0.70, -0.13]) but neither in the text domain (W = 265.00, p = .255; r = 0.20, $CI_{95\%}$=[-0.15, 0.51]) nor the 3D model domain (W = 230.50, p = .242; r = 0.22, $CI_{95\%}$=[-0.14, 0.53]). Regarding the number of ranking interactions to express the preference in an iteration, we found experienced participants express significantly more than novices in text summarization (W = 8439.50, p = .062; r = -0.12, $CI_{95\%}$=[-0.25, 0.02]) but not in photo (W = 19405.50, p = .590; r = -0.03, $CI_{95\%}$=[-0.14, 0.09]) and 3D model (W = 14994.50, p = .516; r = 0.03, $CI_{95\%}$=[-0.09, 0.16]) domains. These results show that \emph{experienced participants express their ranking preference more clearly}. In contrast, \emph{novices might not know if the machine outcome may not be good enough for them, resulting in more incomplete and fewer indifferent preferences}.

\subsubsection{Subjective Satisfaction}

\begin{figure}[t]
\centering
\includegraphics[width=\columnwidth]{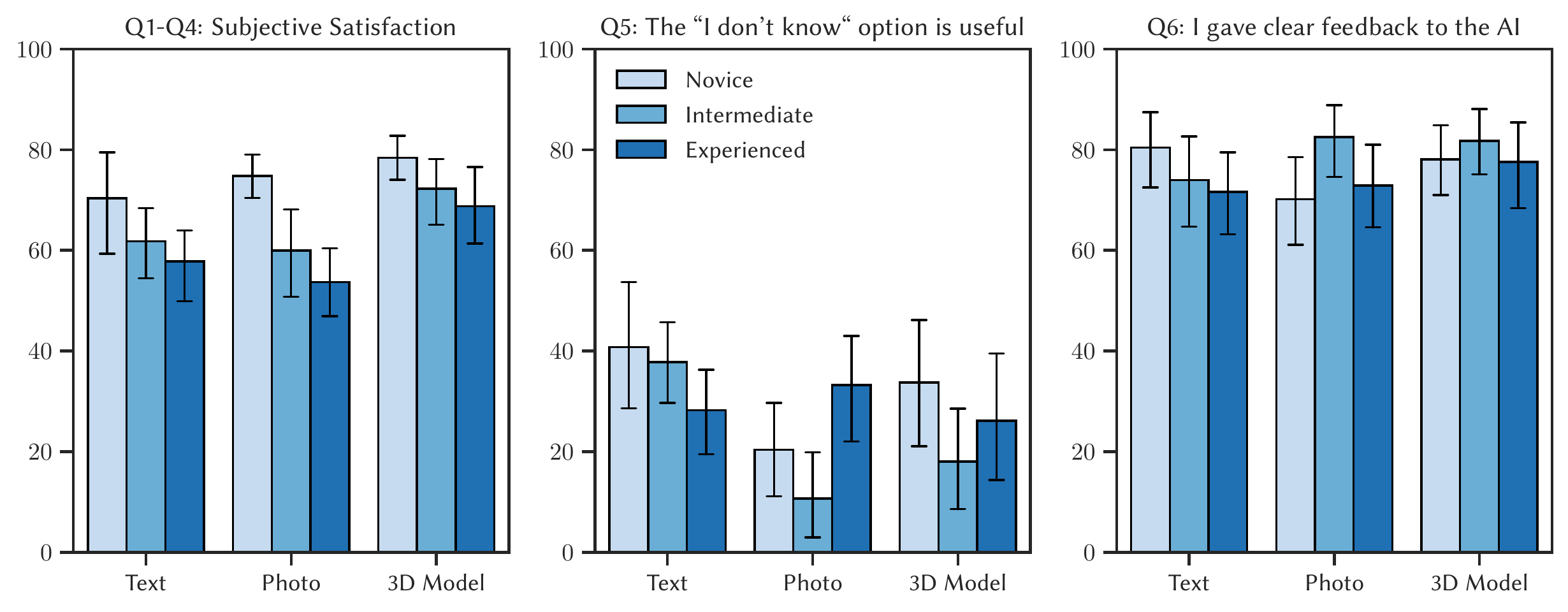}
\caption{Measured subjective satisfaction, the usefulness of providing incomplete preference option while doing the ranking evaluation. The results suggest that subjective satisfaction significantly decreases when comparing novice and experienced participants. All participants considered allowing expressing incomplete preference less useful, and they gave clear feedback to the AI.}
\label{fig:questionnaires}
\end{figure}

As mentioned in~\autoref{sec:satisfy}, we measured subjective satisfaction at the end of every task, and from Q1 to Q4, are used to measure the satisfaction. Since Cronbach's $\alpha$ is fairly high $\alpha$=0.721, $CI_{95\%}$=[0.648, 0.782] in our collected data, we aggregate these questions as satisfaction indicators. See~\autoref{fig:questionnaires}.

We conducted an ART ANOVA~\cite{wobbrock2011art}, as the Shapiro-Wilk normality test showed that the data are not normally distributed (W=.964, p<.001). This analysis revealed that \emph{the overall satisfaction of the final system outcome is significantly influenced by the involved expertise} ($F_{2, 51}$=7.56, p=.001, $\eta^2$=0.23) as well as by the domain context ($F_{2, 51}$=3.84, p=.027, $\eta^2$=0.13). Moreover, no interaction effect was found ($F_{4, 51}$=0.50, p=.733, $\eta^2$=0.04).

\subsection{Interactions within the Optimization Loop}

We analyze three aspects to quantify the overall optimization loop: 1) The directly measured preference utility, i.e., ranking data, from participants. 2) The learned latent utility of the underlying BO optimizer, and 3) The system outcome quality based on objective metrics.
For the directly measured preference utility, a higher value of utility represents participants considering the outcome quality is better in the current evaluating options. The learned latent utility represents how the underlying algorithms consider the human is satisfied with the current results based on the ranking responses; a higher value represents BO optimizer considers more satisfaction on the human side. Lastly, the objectively measured outcome quality metrics measure how different an outcome is from the original task input.

\subsubsection{Measured and Learned Preference Ranking Utility}

\begin{figure*}[t]
    \centering
    \begin{subfigure}[b]{0.33\textwidth}
    \includegraphics[width=\textwidth]{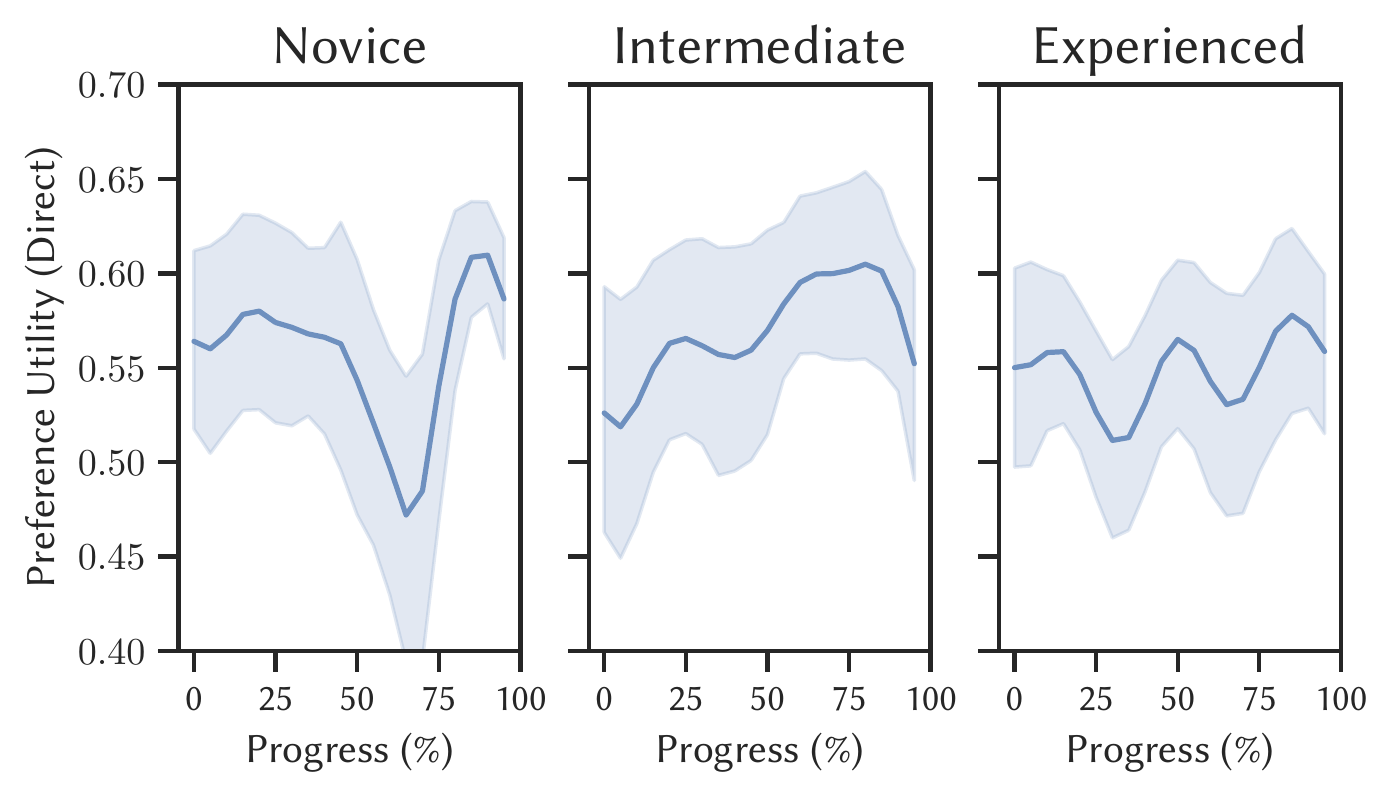}
    \caption{Text Summarization}
    \end{subfigure}
    \hfill
    \begin{subfigure}[b]{0.33\textwidth}
    \includegraphics[width=\textwidth]{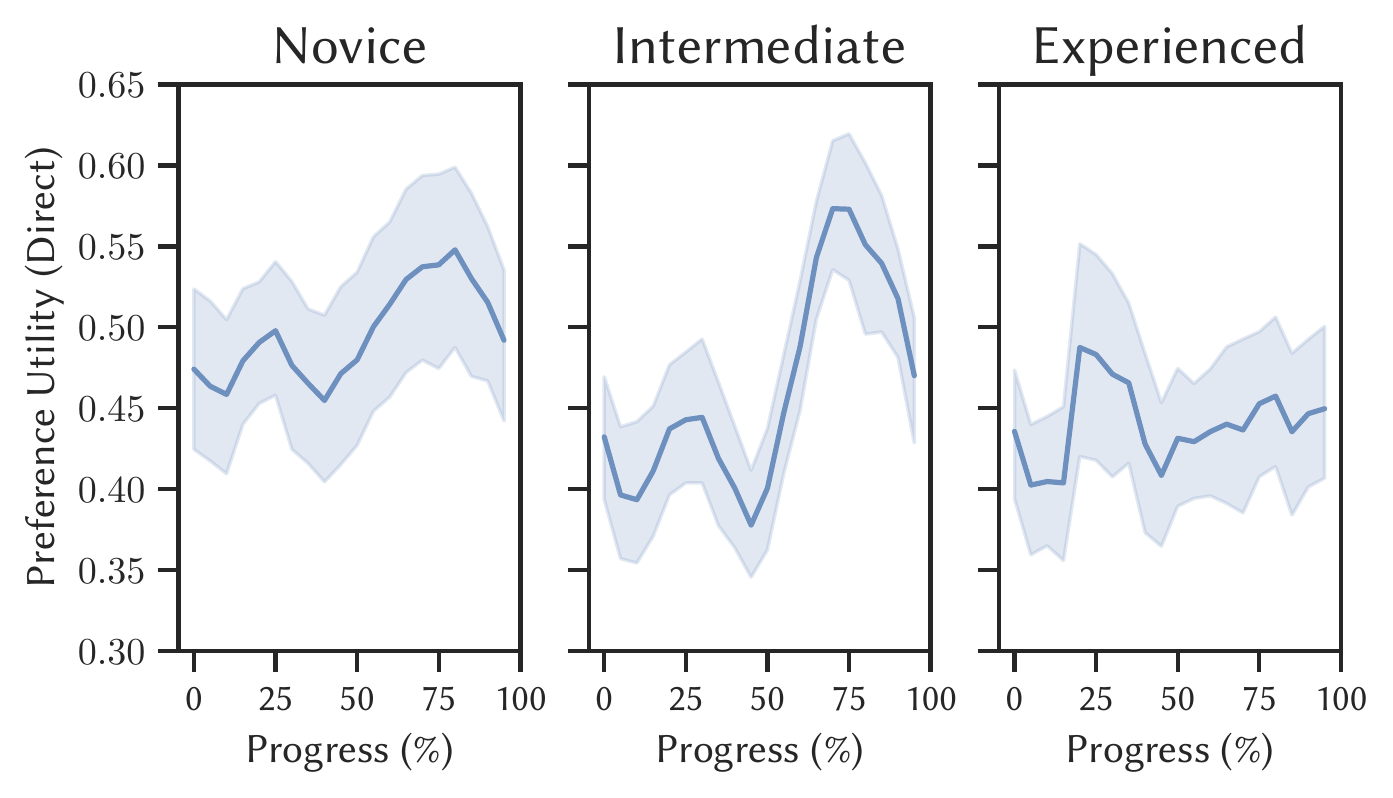}
    \caption{Photo Color Enhancement}
    \end{subfigure}
    \hfill
    \begin{subfigure}[b]{0.33\textwidth}
    \includegraphics[width=\textwidth]{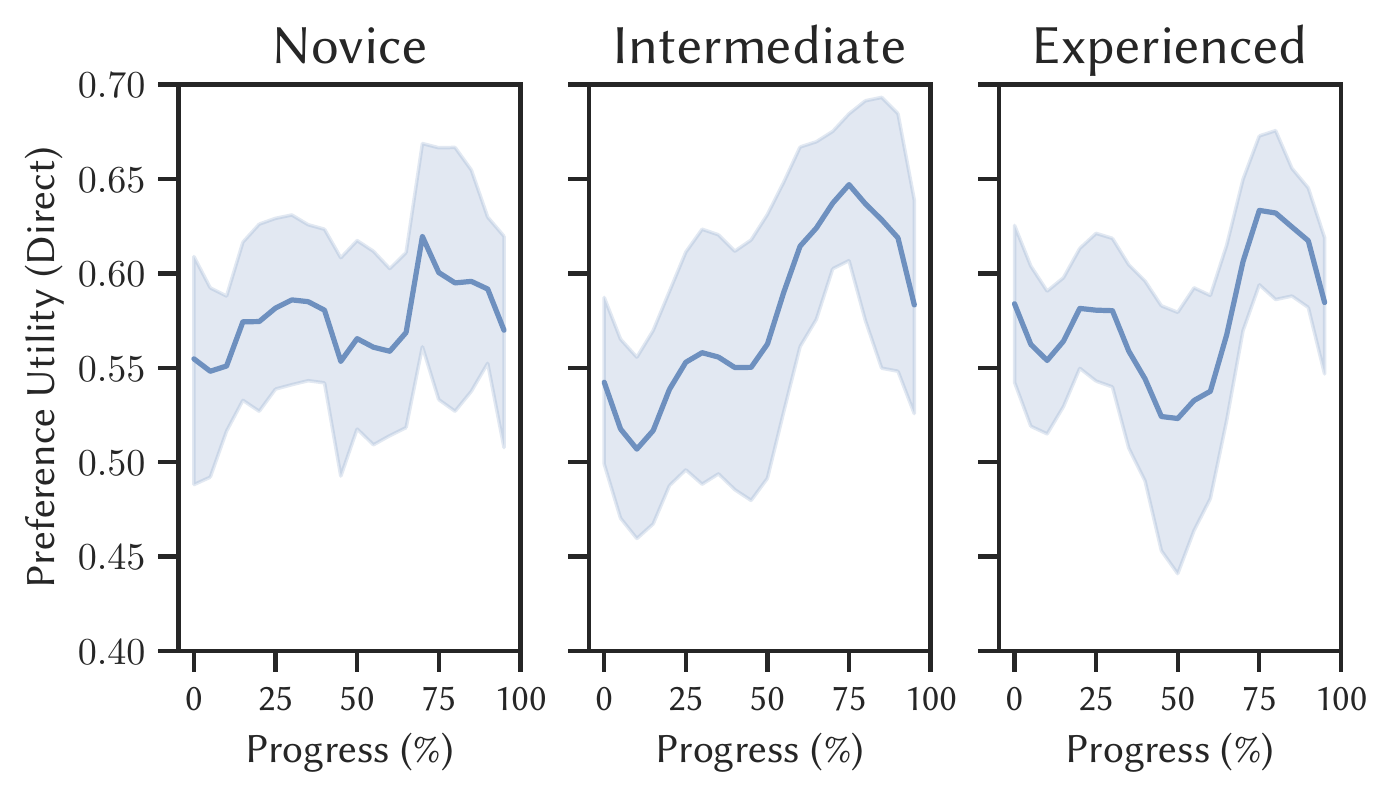}
    \caption{3D Model Simplification}
\end{subfigure}

\caption{Directly measured preference utility: The utility values are normalized from rating labels (Terrible to Excellent). The results indicate that regardless of the involved expertise, participants behave consistently, and in later iterations, the final ranking utility is higher than at the beginning of HITL optimization}
\label{fig:utility-direct}
\end{figure*}

As shown in~\autoref{fig:utility-direct}, for directly measured preference utility from ranking data,
we fitted a linear mixed model~\cite{bates2014fitting, kuznetsova2017lmertest} (estimated using REML and nloptwrap optimizer)
to predict preference utility with involved expertise and exploration iterations. The model included participants as a random effect. Comparing to novice participants ($CI_{95\%}$=[0.49, 0.56], t(3592) = 28.56, p < .001), we found that in all domain contexts, the submitted preference utility from experienced participants is statistically non-significant and
negative ($\beta$ = -0.02, $CI_{95\%}$=[-0.07, 0.03], t(3592) = -0.69, p = .489). The effect of iteration is statistically significant and positive ($\beta$ =
.002, $CI_{95\%}$=[.001, .003], t(3592) = 3.32, p < .001).
This means that regardless of the involved expertise, participants behave consistently, and in later iterations, the final ranking utility is higher than at the beginning of HITL optimization.

\begin{figure*}[t]
    \centering
    \begin{subfigure}[b]{0.33\textwidth}
    \includegraphics[width=\textwidth]{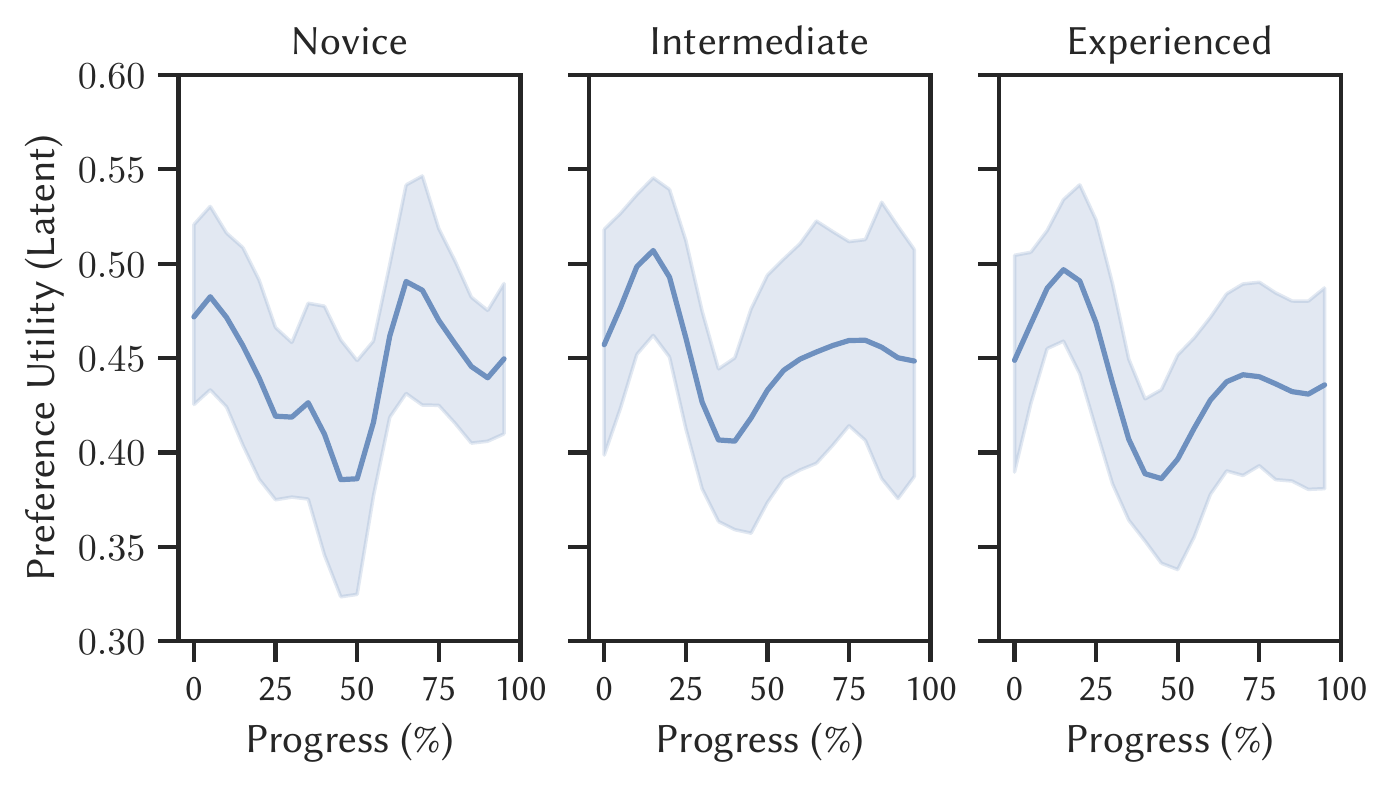}
    \caption{Text Summarization}
    \end{subfigure}
    \hfill
    \begin{subfigure}[b]{0.33\textwidth}
    \includegraphics[width=\textwidth]{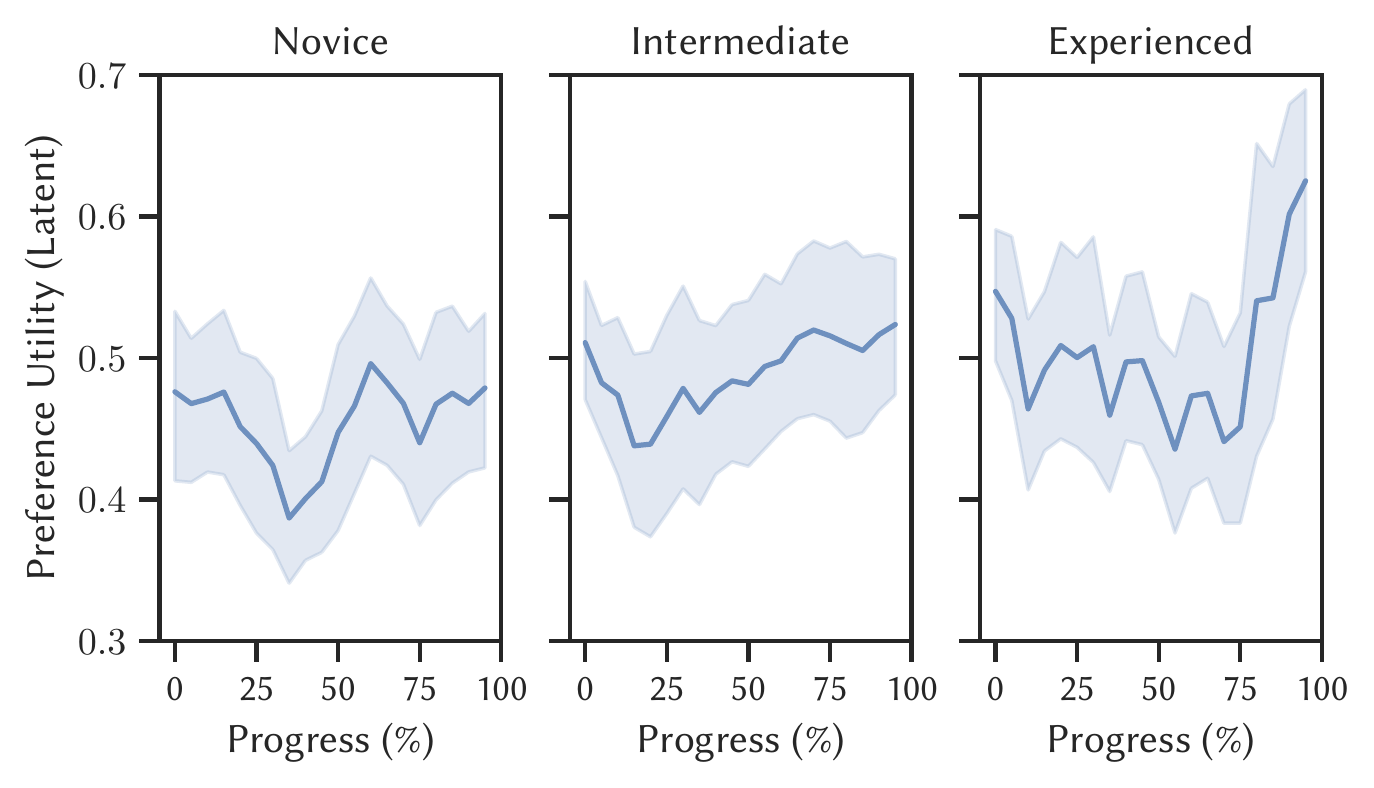}
    \caption{Photo Color Enhancement}
    \end{subfigure}
    \hfill
    \begin{subfigure}[b]{0.33\textwidth}
    \includegraphics[width=\textwidth]{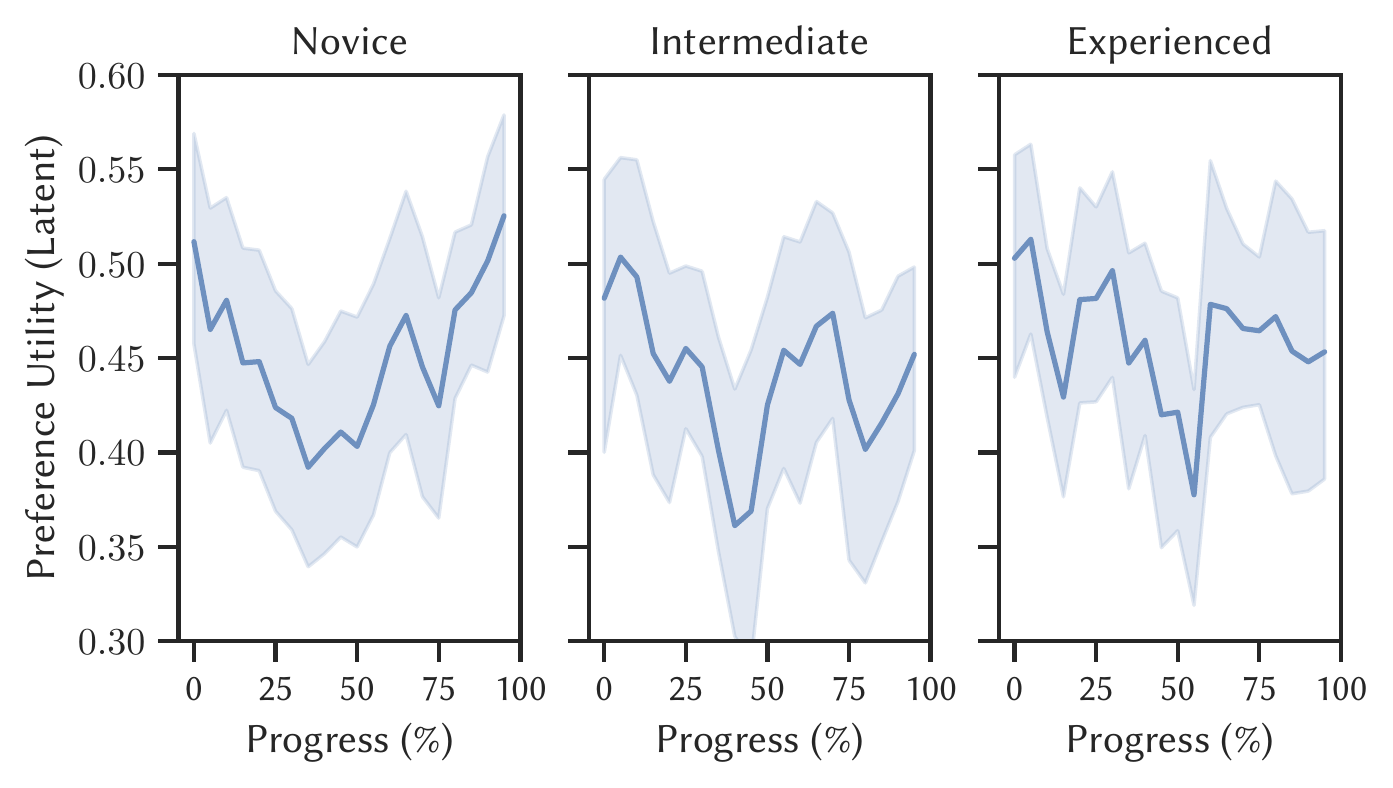}
    \caption{3D Model Simplification}
    \end{subfigure}
    \caption{Learned latent preference utility: The inferred utility values from the machine side (i.e., Bayesian optimization). Our results indicate that provided ranking data from experienced participants are more consistent and more effective in the learning process for the BO optimizer to learn than the ranking data from novices.}
    \label{fig:utility-latent}
\end{figure*}

Regarding the learned latent utility from the BO optimizer, as illustrated in~\autoref{fig:utility-latent}, we fitted another linear mixed model (estimated using REML and nloptwrap optimizer) to predict the learned latent utility with involved expertise and exploration iterations. Comparing to novice participants ($CI_{95\%}$=[0.42, 0.46], t(3592) = 42.97, p < .001), the effect of experienced participants is statistically significant and positive ($\beta$ = 0.03,  $CI_{95\%}$=[0.001, 0.06], t(3592) = 2.03, p = .042). But the effect of iteration is statistically non-significant and positive ($\beta$ = 0.001, $CI_{95\%}$=[-0.0004, 0.002], t(3592) = 1.32, p = .186).
This result means that the provided ranking data from experienced participants are more effective and consistent for the BO optimizer than the ranking data from novices.

\subsubsection{Objective Outcome Quality}

\begin{figure*}[t]
    \centering
    \begin{subfigure}[b]{0.33\textwidth}
    \includegraphics[width=\textwidth]{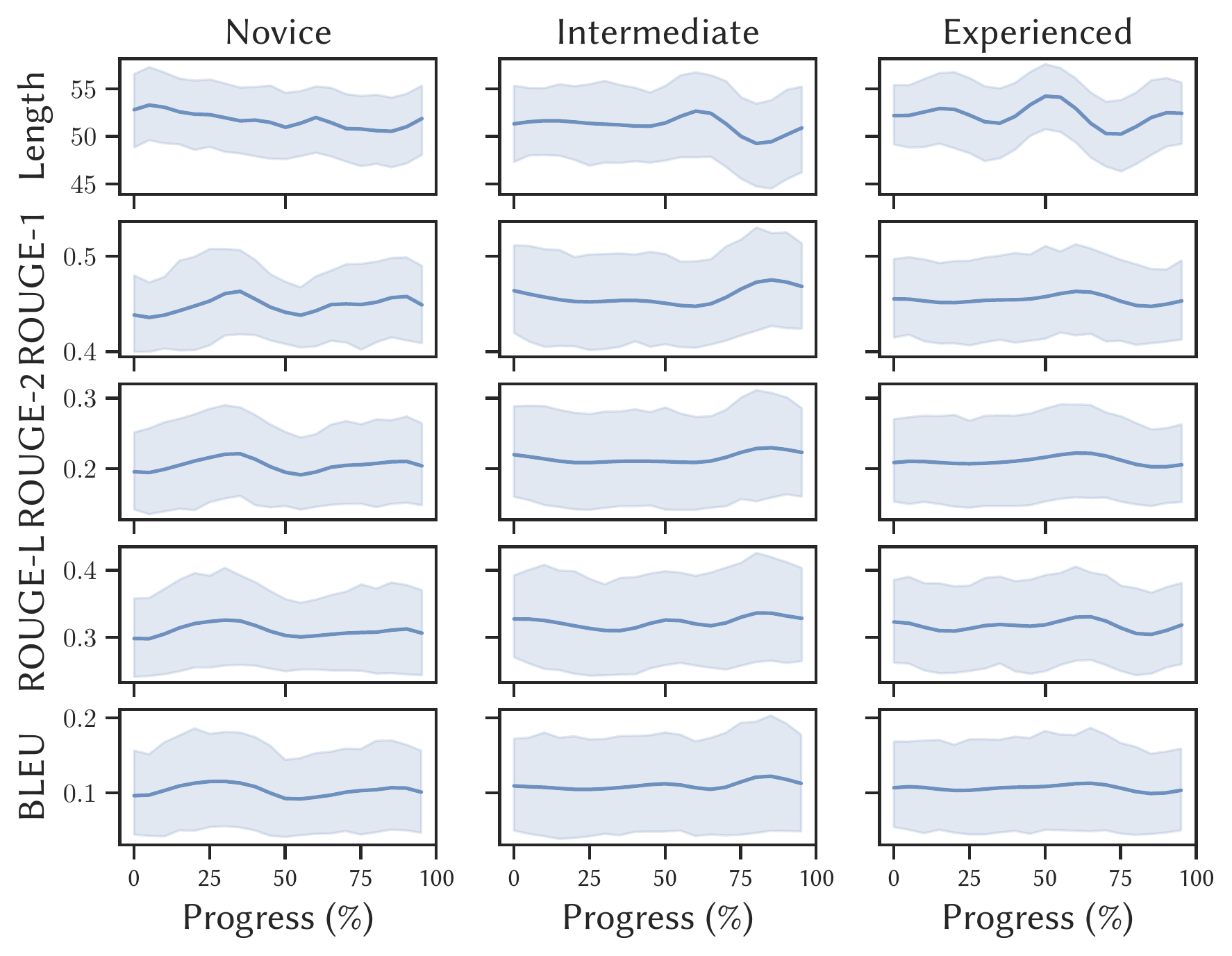}
    \caption{Text Summarization}
    \end{subfigure}
    \hfill
    \begin{subfigure}[b]{0.33\textwidth}
    \includegraphics[width=\textwidth]{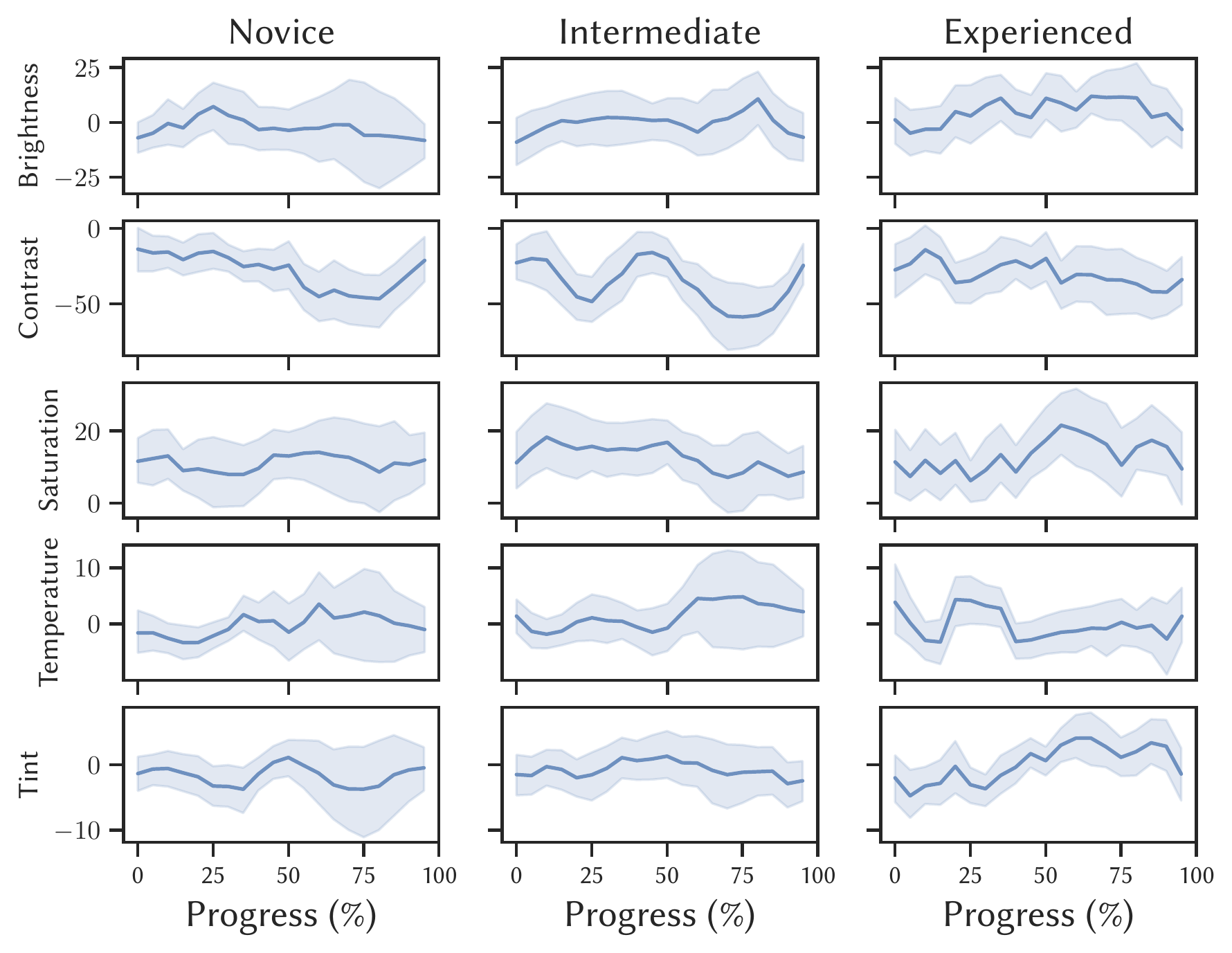}
    \caption{Photo Color Enhancement}
    \end{subfigure}
    \hfill
    \begin{subfigure}[b]{0.33\textwidth}
    \includegraphics[width=\textwidth]{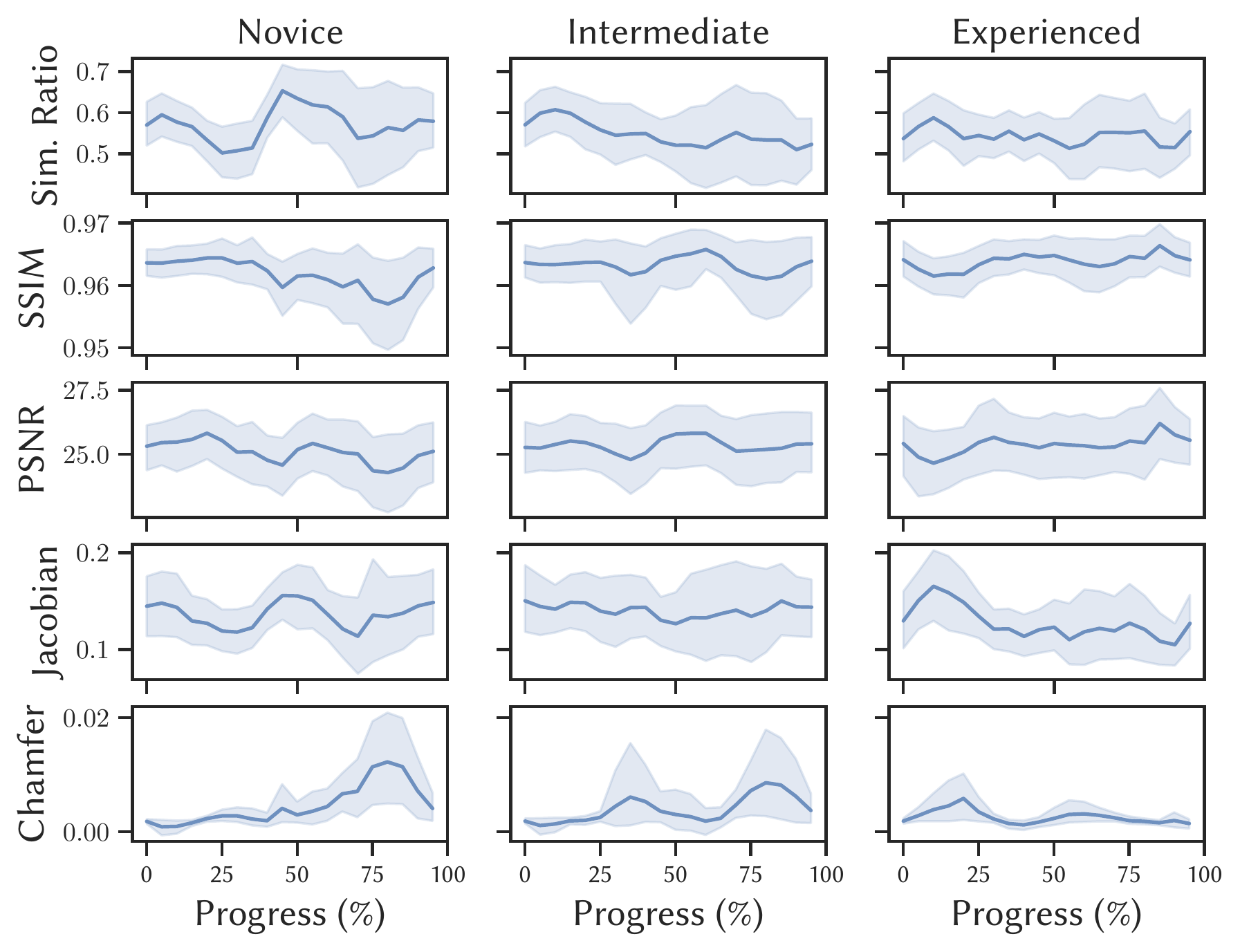}
    \caption{3D Model Simplification}
    \end{subfigure}
    \caption{Objective Quality of System Outcomes: Each context used five metrics to measure the outcome. Experts can identify the technical difference compared to novices, such as minimizing Chamfer distance in 3D model simplification. Overall, our results indicate that novices produce expert-level performance in objective quality.}
    \label{fig:quality-metrics}
\end{figure*}

We normalized the iteration sequence and visualized the exploration progress in~\autoref{fig:quality-metrics}. For analyzing the progress, we fitted linear mixed models for all metrics in the text summarization domain. For example, for length metric: comparing to the results produced by novices ($CI_{95\%}$=[51.20, 54.43], t(1192) = 64.14) is as good as the outcome produced by experienced participants ($\beta$ = -0.20, $CI_{95\%}$=[-2.48, 2.08], t(1192) = -0.17, p = .864), and there are no effects on the involved iteration ($\beta$ = -0.12, $CI_{95\%}$=[-0.26, 0.03], t(1192) = -1.55, p = .120). These results hold the same as for other metrics. In summary, when comparing to outcomes produced when engaging with novices, the effects of involving experienced participants were statistically non-significant, and the effect of iteration was statistically non-significant and negative. This means novices achieved the same level of performance as experienced participants did. These results hold for all metrics we used for measuring outcome quality.

In the photo color enhancement, except for the contracts ($\beta$ = -1.45, $CI_{95\%}$=[-2.05, -0.84], t(1192) = -4.71, p < .001) and temperature ($\beta$ = 0.19, $CI_{95\%}$=[0.004, 0.37], t(1192) = 2.00, p = .045) which are significantly influenced regarding exploration iterations.
The effect on brightness using experienced participants is statistically non-significant and
positive ($\beta$ = 0.76, $CI_{95\%}$=[-10.38, 11.89], t(1192) = 0.13, p = .894) and the effect of iteration is statistically non-significant and negative ($\beta$ =
-0.32, $CI_{95\%}$=[-0.78, 0.14], t(1192) = -1.37, p = .172), when compared to novices ($CI_{95\%}$=[-7.11, 8.02], t(1192) = 0.12, p = .907), and these results are the same for saturation and tint metrics.

Lastly, for 3D model simplification, we found that experienced participants ($\beta$ = 0.003, $CI_{95\%}$=[0.0001, 0.007], t(1192) = 2.00, p = .046) outperformed novices ($CI_{95\%}$=[-0.002, 0.002], t(1192) = 0.10) only in keeping surface distance low, meaning better in maintaining surface quality. We did not find significant differences in other metrics when comparing experienced users' and novices' outcomes. This result means that experienced participants are better at identifying technical differences as surface quality is less observable, as discussed in~\autoref{sec:metrics}. However, novices can achieve expert-level performance under the HITL optimization context, similar to other contexts.

\section{Discussion and Implications}

The results in \autoref{sec:results} could be summarized into two major observations: 1)~Novices can achieve expert-level performance in objective quality in all cases. 2)~Participants with higher expertise show more explicit preferences, dissatisfaction, and iterations, but novices are more quickly terminated and show more satisfaction. Below, we will discuss what implications we think these observations might have.

\subsection{Outcome Quality and Pareto Optimality}

When we have a well-defined metric that can measure the quality of an outcome, the optimization process could be done procedurally using a machine alone. However, in reality, the outcome quality is often characterized by a set of metrics, and \emph{Pareto optimality}~\cite{pareto1912manuel} is a useful concept for discussing machine rationality regarding its outcome quality. \emph{Pareto optimality} describes a trade-off situation where a system outcome is optimal if any improvements in one objective result in the deterioration of others. This trade-off is also called the \emph{Pareto front}, and outcomes on this front refer to \emph{Pareto frontiers}.
Conceptually, the Pareto optimality captures the measurable components when evaluating an outcome, whereas non-measurable parts reflect more about the subjective matter. Let $\mathcal{P}_{s}$ be the system parameter space defined by $[0, 1]^{r} (s\in \mathbf{N}^{+})$, and $\mathcal{O}$ be the outcome space generated from the parameter space. Then, the rational component of a HITL optimization is to explore the outcome space $\mathcal{O}$ concerning a given set of objective metrics $\mathcal{M}_{t} (t\in\mathbf{N}^{+})$. The Pareto front $\mathcal{F}$ is determined by the outcome space and specified metrics, which essentially depends on the parameter space and metrics, i.e., $\mathcal{F} (\mathcal{P}_{s}, \mathcal{M}_{t})$, which captures the boundary of machine rationality and HITL optimization could be considered as the exploration in this space to reach the Pareto front.

This concept avoids the aggregation problem of contradicted multi-objective objectives, such as in our user tasks, participants need to summarize the text while preserving the meaning or simplify 3D models as much as possible while keeping the overall appearance.
However, note that converging to the true Pareto optimal set has a technical challenge, and yet still in active research~\cite{daulton2021parallel, schulz2018interactive}, as there might be an infinite amount of candidates, and metrics might interact with each other. Instead of evaluating whether an outcome is a Pareto frontier, it is more useful to discuss whether the optimization made any progress to guarantee the final outcome is more dominant than the initial ones.

In our results, we showed that both novices and experienced participants improved the objective measures and could achieve a similar level of quality, meaning the final outcomes are Pareto dominant than the initial ones.
Under the Pareto optimality framework, the BO learns the underlying preference using users' ranking choices, which tend to converge to different non-Pareto optimal results. But since the BO optimizer assumes human has a stable preference utility function that will eventually converge, we argue that novice participants do not have enough evaluation metrics in mind, and the system outcome does not necessarily need to arrive at the front. In contrast, experts attempt to keep optimizing or exploring other objectives when machine rationality already reaches the objective Pareto front.
Hence, compared to experienced participants who potentially evaluate more metrics than the machine, more flaws might be discovered in this process, and cause either more uncertain in expressing its decision and causing more decision time (e.g., in text summarization) or easier to form a decision and cause less decision time (e.g., in the photo and 3D model contexts). Since experts report significantly higher dissatisfaction than novices, we argue that this result shows a mismatch of the Pareto front between the participants and machine rationality, and the source of the dissatisfaction comes from the involved expertise.

\subsection{Expertise and Satisficing Decision Strategy}

Based on the analysis of the outcome quality from the HITL optimization loop, we did not find enough evidence to indicate a significant difference regarding the quality of the system outcome between different levels of expertise. However, with increasing expertise, overall user satisfaction decreases, and the number of iterations increases. This observed behavior matches the maximizing decision strategy since participants are asked to terminate at satisfaction, and experts attempt to explore the solution space significantly more than novices. Since the involved expertise is increased, more flaws in the system may be discovered in this process, resulting in more dissatisfaction. This observation suggests that we could involve more expertise to identify more system flaws iteratively while exploring the solution space. Although machine rationality would not be improved without a reparameterization of the underlying algorithm, this observed behavior could be used as an indicator in hindsight analysis to inform system designers to 1) improve underlying machine rationality, 2) further improve the HITL optimization process, and 3) better support users to explore desired solutions. For novices, using a satisficing strategy is good enough to get to expert-level performance with the help of HITL optimization.

\subsection{The Impact of Involved Expertise}

The objective outcome quality might not depend on the involved human expertise when a machine learner baked enough domain knowledge in its underlying algorithm. What might be the ``minimum'' required expertise to obtain meaningful machine outputs, then? What if a user constantly provides flawed random choices? Intuitively, such a condition would not benefit a preference-optimized HITL system.
Admittedly, to evaluate the behavior between ``zero expertise'' and ``novice,'' we could program a random choice generator to test and observe the results. Still, we are bound to a limited observation time and two implicit assumptions. The first assumption is that the expertise level has a total order, and a random choice generator is a minimum element for all levels of expertise. Second, a random choice generator can never produce a meaningful outcome in the context of HITL optimization.

These two assumptions might be considered true at first sight. However, we cannot compare the amount of expertise from a random choice generator or an intelligent human being. Notably, the Borel–Cantelli lemma~\cite{cohn2013measure}\footnote{In proposition 10.2.2 (b).},
states that with an infinite number of events, the probability\footnote{Strictly speaking, the event happens \emph{almost surely} as the Lebesgue measure is 1.}
of observing a meaningful result is 1. This theory explains that even with a random choice generator, as long as it continues to generate choices, a meaningful sequence of choices eventually will occur, such that the HITL system can produce desired outcomes.
In other words, this theoretical fact endorses that a sufficient amount of expertise could be beneficial to produce meaningful outcomes in a short amount of time comparably, and our results complement that more involved expertise creates increased iterations of interactions for explorations.

\subsection{Limitations and Future Work}
Although we allowed users to express ``I don't know'' as their incomplete preference, a participant may still provide a sub-optimal ranking due to fatigue from a long time of participation or other relevant reasons, resulting in the violation of the incompleteness assumption. From an algorithmic perspective, although the PBO handles ideal randomized choices, the provided ranking choices might even be worse than assumed Gaussian distributed random choices due to subjective reasons. Besides, the underlying preferences might change at every iteration. For example, experts may further reason for using the system outcomes or trying to make sense of the sequential outcomes. Instead, novice users judge locally, making their behavior much more stable. The choice of objective quality evaluation metrics may also impact the interpretation of the optimization process due to their interaction effect.

One of the conventional motivations for developing an objective metric is to use it to predict human judgments. The development of an objective metric implicitly assumes common sense among the crowd, and the metric may not be suitable for measuring individual preferences. Instead of asking users for their judgment to explore the solution space, it might be more interesting for future research to utilize human judgment more in exploring dynamic solution spaces where human is only involved when the machine reaches its boundary of rationality.
Furthermore, instead of evaluating the impact of expertise on the exploration behavior of one static solution space, we could evaluate the interaction effect of the involved expertise and the underlying HITL optimizer. For instance, one could design an experiment to understand the decision behavior on the Pareto front where all machine-proposed options are objectively optimal. It would be interesting to check how the involved expertise impacts the decision behavior among all objectively optimal Pareto frontiers and, thus, better understand the difference between subjective and objective Pareto fronts.

\section{Summary and Outlook}
In this paper, we evaluated three example contexts to understand the impact of involving different levels of human expertise in HITL optimization on the subjective satisfaction of system outcomes quality. Our study answered our initial research questions:
\textbf{RQ1}~using a human in the loop to optimize system outcomes allows novice users to achieve expert-level performance;
\textbf{RQ2}~with decreasing expertise, the eventual subjective satisfaction increases, and the entire process terminates faster.

Our findings contradict the intuition that using higher expertise leads to better results. Instead, when collaborating with a machine learner, users without a sufficient amount of domain expertise can still show a compatible level of performance as experts, with even higher satisfaction. We argue interpretations of these observations:
1)~When humans do not have enough insights to evaluate the quality of the system outcomes, the eventual result reflects the performance of the machine algorithm.
2)~Expert users express less satisfaction when having more insights than the underlying algorithm, and the underlying machine rationality limits the outcome of the optimization loop. In this case, the satisfaction of a human can be used as an indicator to inform system designers further to improve their underlying machine algorithm. The insights suggest optimization using human feedback may be more helpful and favor exploration purposes rather than using them for exploiting the solution space covered by machine rationality. Our results bring us closer to better human models and system design principles for exploiting human intelligence. Inferring and adapting to user expertise also play a pivotal role in achieving successful interaction. An intelligent machine system that adequately considers human expertise can help users improve their skills and achieve higher user satisfaction. %

\section{Open Science}
\label{sec:open-science}
We encourage readers to reproduce and extend our results. We open-sourced the collected dataset, systems, and analysis scripts at \url{https://changkun.de/s/expertise-loop}.

\begin{acks}
We thank Heiko Drewes for his insights regarding expertise;
Dennis Dietz, Yi Xia, Guoliang Xue, Yifei Zhan, Daniel Buschek, and Francesco Chiossi for their inspiring discussions regarding user satisfaction, preference, choice, and decision-making;
Eyke Hüllermeier and Karlson Pfannschmidt for the useful discussions of context-depended ranking.
\end{acks}

\bibliographystyle{ACM-Reference-Format}
\bibliography{ref}


\begin{thebibliography}{72}


\ifx \showCODEN    \undefined \def \showCODEN     #1{\unskip}     \fi
\ifx \showDOI      \undefined \def \showDOI       #1{#1}\fi
\ifx \showISBNx    \undefined \def \showISBNx     #1{\unskip}     \fi
\ifx \showISBNxiii \undefined \def \showISBNxiii  #1{\unskip}     \fi
\ifx \showISSN     \undefined \def \showISSN      #1{\unskip}     \fi
\ifx \showLCCN     \undefined \def \showLCCN      #1{\unskip}     \fi
\ifx \shownote     \undefined \def \shownote      #1{#1}          \fi
\ifx \showarticletitle \undefined \def \showarticletitle #1{#1}   \fi
\ifx \showURL      \undefined \def \showURL       {\relax}        \fi
\providecommand\bibfield[2]{#2}
\providecommand\bibinfo[2]{#2}
\providecommand\natexlab[1]{#1}
\providecommand\showeprint[2][]{arXiv:#2}

\bibitem[Anand(1987)]%
        {anand1987axioms}
\bibfield{author}{\bibinfo{person}{Paul Anand}.}
  \bibinfo{year}{1987}\natexlab{}.
\newblock \showarticletitle{Are the preference axioms really rational?}
\newblock \bibinfo{journal}{\emph{Theory and Decision}} \bibinfo{volume}{23},
  \bibinfo{number}{2} (\bibinfo{date}{Sep} \bibinfo{year}{1987}),
  \bibinfo{pages}{189–214}.
\newblock
\showISSN{1573-7187}
\urldef\tempurl%
\url{https://doi.org/10.1007/BF00126305}
\showDOI{\tempurl}


\bibitem[Arvo et~al\mbox{.}(2015)]%
        {arvo2015survey}
\bibfield{author}{\bibinfo{person}{Jukka Arvo}, \bibinfo{person}{Antti
  Euranto}, \bibinfo{person}{Lauri Jarvenpaa}, \bibinfo{person}{Teijo
  Lehtonen}, {and} \bibinfo{person}{Timo Knuutila}.}
  \bibinfo{year}{2015}\natexlab{}.
\newblock \bibinfo{booktitle}{\emph{3D Mesh Simplification – A survey of
  algorithms and CAD model simplification tests}}.
\newblock \bibinfo{publisher}{University of Turku}, \bibinfo{address}{Turku,
  Finland}.
\newblock
\urldef\tempurl%
\url{http://urn.fi/URN:ISBN:978-951-29-6202-0}
\showURL{%
\tempurl}


\bibitem[Balandat et~al\mbox{.}(2020)]%
        {balandat2020botorch}
\bibfield{author}{\bibinfo{person}{Maximilian Balandat}, \bibinfo{person}{Brian
  Karrer}, \bibinfo{person}{Daniel~R. Jiang}, \bibinfo{person}{Samuel Daulton},
  \bibinfo{person}{Benjamin Letham}, \bibinfo{person}{Andrew~Gordon Wilson},
  {and} \bibinfo{person}{Eytan Bakshy}.} \bibinfo{year}{2020}\natexlab{}.
\newblock \showarticletitle{BOTORCH: A Framework for Efficient Monte-Carlo
  Bayesian Optimization}. In \bibinfo{booktitle}{\emph{Proceedings of the 34th
  International Conference on Neural Information Processing Systems}}
  (Vancouver, BC, Canada) \emph{(\bibinfo{series}{NIPS'20})}.
  \bibinfo{publisher}{Curran Associates Inc.}, \bibinfo{address}{Red Hook, NY,
  USA}, Article \bibinfo{articleno}{1807}, \bibinfo{numpages}{15}~pages.
\newblock
\showISBNx{9781713829546}


\bibitem[Bansal et~al\mbox{.}(2019)]%
        {bansal2019dms}
\bibfield{author}{\bibinfo{person}{Gagan Bansal}, \bibinfo{person}{Besmira
  Nushi}, \bibinfo{person}{Ece Kamar}, \bibinfo{person}{Daniel~S. Weld},
  \bibinfo{person}{Walter~S. Lasecki}, {and} \bibinfo{person}{Eric Horvitz}.}
  \bibinfo{year}{2019}\natexlab{}.
\newblock \showarticletitle{Updates in Human-AI Teams: Understanding and
  Addressing the Performance/Compatibility Tradeoff}.
\newblock \bibinfo{journal}{\emph{Proceedings of the AAAI Conference on
  Artificial Intelligence}} \bibinfo{volume}{33}, \bibinfo{number}{0101}
  (\bibinfo{date}{Jul} \bibinfo{year}{2019}), \bibinfo{pages}{2429–2437}.
\newblock
\showISSN{2374-3468}
\urldef\tempurl%
\url{https://doi.org/10.1609/aaai.v33i01.33012429}
\showDOI{\tempurl}


\bibitem[Barrow et~al\mbox{.}(1977)]%
        {barrow1977chamfer}
\bibfield{author}{\bibinfo{person}{H.~G. Barrow}, \bibinfo{person}{J.~M.
  Tenenbaum}, \bibinfo{person}{R.~C. Bolles}, {and} \bibinfo{person}{H.~C.
  Wolf}.} \bibinfo{year}{1977}\natexlab{}.
\newblock \bibinfo{title}{Parametric Correspondence and Chamfer Matching: Two
  New Techniques for Image Matching}.
\newblock
\newblock
\urldef\tempurl%
\url{https://apps.dtic.mil/sti/citations/ADA458355}
\showURL{%
\tempurl}


\bibitem[Bates et~al\mbox{.}(2014)]%
        {bates2014fitting}
\bibfield{author}{\bibinfo{person}{Douglas Bates}, \bibinfo{person}{Martin
  Mächler}, \bibinfo{person}{Ben Bolker}, {and} \bibinfo{person}{Steve
  Walker}.} \bibinfo{year}{2014}\natexlab{}.
\newblock \bibinfo{title}{Fitting Linear Mixed-Effects Models using lme4}.
\newblock
\newblock
\urldef\tempurl%
\url{https://doi.org/10.48550/ARXIV.1406.5823}
\showDOI{\tempurl}


\bibitem[Benavoli et~al\mbox{.}(2021)]%
        {benavoli2021choice}
\bibfield{author}{\bibinfo{person}{Alessio Benavoli}, \bibinfo{person}{Dario
  Azzimonti}, {and} \bibinfo{person}{Dario Piga}.}
  \bibinfo{year}{2021}\natexlab{}.
\newblock \bibinfo{title}{Choice functions based multi-objective Bayesian
  optimisation}.
\newblock
\newblock
\urldef\tempurl%
\url{https://doi.org/10.48550/arXiv.2110.08217}
\showDOI{\tempurl}


\bibitem[Botsch et~al\mbox{.}(2010)]%
        {botsch2010polygon}
\bibfield{author}{\bibinfo{person}{Mario Botsch}, \bibinfo{person}{Leif
  Kobbelt}, \bibinfo{person}{Mark Pauly}, \bibinfo{person}{Pierre Alliez},
  {and} \bibinfo{person}{Bruno L{\'e}vy}.} \bibinfo{year}{2010}\natexlab{}.
\newblock \bibinfo{booktitle}{\emph{Polygon mesh processing}}.
\newblock \bibinfo{publisher}{CRC press}, \bibinfo{address}{USA}.
\newblock


\bibitem[Bourne et~al\mbox{.}(2014)]%
        {bourne2014def}
\bibfield{author}{\bibinfo{person}{Lyle Bourne}, \bibinfo{person}{James Kole},
  {and} \bibinfo{person}{Alice Healy}.} \bibinfo{year}{2014}\natexlab{}.
\newblock \showarticletitle{Expertise: defined, described, explained}.
\newblock \bibinfo{journal}{\emph{Frontiers in Psychology}}
  \bibinfo{volume}{5} (\bibinfo{year}{2014}), \bibinfo{pages}{186}.
\newblock
\showISSN{1664-1078}
\urldef\tempurl%
\url{https://doi.org/10.3389/fpsyg.2014.00186}
\showDOI{\tempurl}


\bibitem[Brandstätter et~al\mbox{.}(2006)]%
        {brandstaetter2006tradeoff}
\bibfield{author}{\bibinfo{person}{Eduard Brandstätter}, \bibinfo{person}{Gerd
  Gigerenzer}, {and} \bibinfo{person}{Ralph Hertwig}.}
  \bibinfo{year}{2006}\natexlab{}.
\newblock \showarticletitle{The Priority Heuristic: Making Choices Without
  Trade-Offs}.
\newblock \bibinfo{journal}{\emph{Psychological review}} \bibinfo{volume}{113},
  \bibinfo{number}{2} (\bibinfo{date}{Apr} \bibinfo{year}{2006}),
  \bibinfo{pages}{409–432}.
\newblock
\showISSN{0033-295X}
\urldef\tempurl%
\url{https://doi.org/10.1037/0033-295X.113.2.409}
\showDOI{\tempurl}


\bibitem[Brochu et~al\mbox{.}(2010)]%
        {brochu2010animation_design}
\bibfield{author}{\bibinfo{person}{Eric Brochu}, \bibinfo{person}{Tyson
  Brochu}, {and} \bibinfo{person}{Nando de Freitas}.}
  \bibinfo{year}{2010}\natexlab{}.
\newblock \showarticletitle{A {Bayesian} Interactive Optimization Approach to
  Procedural Animation Design}. In \bibinfo{booktitle}{\emph{Proceedings of the
  2010 {ACM} {SIGGRAPH}/{Eurographics} {Symposium} on {Computer} {Animation}}}
  (Madrid, Spain) \emph{(\bibinfo{series}{{SCA} '10})}.
  \bibinfo{publisher}{Eurographics Association}, \bibinfo{address}{Goslar,
  DEU}, \bibinfo{pages}{103--112}.
\newblock
\urldef\tempurl%
\url{https://dl.acm.org/doi/10.5555/1921427.1921443}
\showURL{%
\tempurl}


\bibitem[Brochu et~al\mbox{.}(2007)]%
        {brochu2007activelearn}
\bibfield{author}{\bibinfo{person}{Eric Brochu}, \bibinfo{person}{Nando~de
  Freitas}, {and} \bibinfo{person}{Abhijeet Ghosh}.}
  \bibinfo{year}{2007}\natexlab{}.
\newblock \showarticletitle{Active preference learning with discrete choice
  data}. In \bibinfo{booktitle}{\emph{Proceedings of the 20th {International}
  {Conference} on {Neural} {Information} {Processing} {Systems}}}
  \emph{(\bibinfo{series}{{NIPS}'07})}. \bibinfo{publisher}{Curran Associates
  Inc.}, \bibinfo{address}{Red Hook, NY, USA}, \bibinfo{pages}{409--416}.
\newblock
\showISBNx{978-1-60560-352-0}
\urldef\tempurl%
\url{https://dl.acm.org/doi/abs/10.5555/2981562.2981614}
\showURL{%
\tempurl}


\bibitem[Chan et~al\mbox{.}(2022)]%
        {chan2022novice}
\bibfield{author}{\bibinfo{person}{Liwei Chan}, \bibinfo{person}{Yi-Chi Liao},
  \bibinfo{person}{George~B. Mo}, \bibinfo{person}{John~J. Dudley},
  \bibinfo{person}{Chun-Lien Cheng}, \bibinfo{person}{Per~Ola Kristensson},
  {and} \bibinfo{person}{Antti Oulasvirta}.} \bibinfo{year}{2022}\natexlab{}.
\newblock \showarticletitle{Investigating Positive and Negative Qualities of
  Human-in-the-Loop Optimization for Designing Interaction Techniques}. In
  \bibinfo{booktitle}{\emph{Proceedings of the 2022 CHI Conference on Human
  Factors in Computing Systems}} \emph{(\bibinfo{series}{CHI ’22})}.
  \bibinfo{publisher}{Association for Computing Machinery},
  \bibinfo{address}{New York, NY, USA}, \bibinfo{pages}{1--14}.
\newblock
\urldef\tempurl%
\url{https://doi.org/10.1145/3491102.3501850}
\showDOI{\tempurl}


\bibitem[Cohn(2013)]%
        {cohn2013measure}
\bibfield{author}{\bibinfo{person}{Donald~L. Cohn}.}
  \bibinfo{year}{2013}\natexlab{}.
\newblock \bibinfo{booktitle}{\emph{Probability}}.
\newblock \bibinfo{publisher}{Springer}, \bibinfo{address}{New York, NY}.
  307–371 pages.
\newblock
\showISBNx{978-1-4614-6956-8}
\urldef\tempurl%
\url{https://doi.org/10.1007/978-1-4614-6956-8_10}
\showDOI{\tempurl}


\bibitem[Collins(2013)]%
        {collins2013dim3}
\bibfield{author}{\bibinfo{person}{Harry Collins}.}
  \bibinfo{year}{2013}\natexlab{}.
\newblock \showarticletitle{Three dimensions of expertise}.
\newblock \bibinfo{journal}{\emph{Phenomenology and the Cognitive Sciences}}
  \bibinfo{volume}{12}, \bibinfo{number}{2} (\bibinfo{date}{June}
  \bibinfo{year}{2013}), \bibinfo{pages}{253--273}.
\newblock
\showISSN{1572-8676}
\urldef\tempurl%
\url{https://doi.org/10.1007/s11097-011-9203-5}
\showDOI{\tempurl}


\bibitem[Corsini et~al\mbox{.}(2013)]%
        {corsini2013perceptual}
\bibfield{author}{\bibinfo{person}{Massimiliano Corsini},
  \bibinfo{person}{Mohamed-Chaker Larabi}, \bibinfo{person}{Guillaume
  Lavou{\'e}}, \bibinfo{person}{Old{\v{r}}ich Pet{\v{r}}{\'\i}k},
  \bibinfo{person}{Libor V{\'a}{\v{s}}a}, {and} \bibinfo{person}{Kai Wang}.}
  \bibinfo{year}{2013}\natexlab{}.
\newblock \showarticletitle{Perceptual Metrics for Static and Dynamic Triangle
  Meshes}.
\newblock \bibinfo{journal}{\emph{Computer Graphics Forum}}
  \bibinfo{volume}{32}, \bibinfo{number}{1} (\bibinfo{year}{2013}),
  \bibinfo{pages}{101–125}.
\newblock
\urldef\tempurl%
\url{https://doi.org/10.1111/cgf.12001}
\showDOI{\tempurl}


\bibitem[Daulton et~al\mbox{.}(2021)]%
        {daulton2021parallel}
\bibfield{author}{\bibinfo{person}{Samuel Daulton}, \bibinfo{person}{Maximilian
  Balandat}, {and} \bibinfo{person}{Eytan Bakshy}.}
  \bibinfo{year}{2021}\natexlab{}.
\newblock \showarticletitle{Parallel Bayesian Optimization of Multiple Noisy
  Objectives with Expected Hypervolume Improvement}. In
  \bibinfo{booktitle}{\emph{Advances in Neural Information Processing
  Systems}}, \bibfield{editor}{\bibinfo{person}{M.~Ranzato},
  \bibinfo{person}{A.~Beygelzimer}, \bibinfo{person}{Y.~Dauphin},
  \bibinfo{person}{P.S. Liang}, {and} \bibinfo{person}{J.~Wortman Vaughan}}
  (Eds.), Vol.~\bibinfo{volume}{34}. \bibinfo{publisher}{Curran Associates,
  Inc.}, \bibinfo{address}{Red Hook, NY, USA}, \bibinfo{pages}{2187--2200}.
\newblock
\urldef\tempurl%
\url{https://proceedings.neurips.cc/paper/2021/file/11704817e347269b7254e744b5e22dac-Paper.pdf}
\showURL{%
\tempurl}


\bibitem[Dietvorst et~al\mbox{.}(2015)]%
        {dietvorst2015algaversion}
\bibfield{author}{\bibinfo{person}{Berkeley~J. Dietvorst},
  \bibinfo{person}{Joseph~P. Simmons}, {and} \bibinfo{person}{Cade Massey}.}
  \bibinfo{year}{2015}\natexlab{}.
\newblock \showarticletitle{Algorithm aversion: people erroneously avoid
  algorithms after seeing them err}.
\newblock \bibinfo{journal}{\emph{Journal of Experimental Psychology. General}}
  \bibinfo{volume}{144}, \bibinfo{number}{1} (\bibinfo{date}{Feb}
  \bibinfo{year}{2015}), \bibinfo{pages}{114–126}.
\newblock
\showISSN{1939-2222}
\urldef\tempurl%
\url{https://doi.org/10.1037/xge0000033}
\showDOI{\tempurl}


\bibitem[Ehsan et~al\mbox{.}(2021)]%
        {ehsan2021explain}
\bibfield{author}{\bibinfo{person}{Upol Ehsan}, \bibinfo{person}{Q.~Vera Liao},
  \bibinfo{person}{Michael Muller}, \bibinfo{person}{Mark~O. Riedl}, {and}
  \bibinfo{person}{Justin~D. Weisz}.} \bibinfo{year}{2021}\natexlab{}.
\newblock \showarticletitle{Expanding Explainability: Towards Social
  Transparency in AI systems}. In \bibinfo{booktitle}{\emph{Proceedings of the
  2021 CHI Conference on Human Factors in Computing Systems}}
  \emph{(\bibinfo{series}{CHI'21})}. \bibinfo{publisher}{Association for
  Computing Machinery}, \bibinfo{address}{New York, NY, USA},
  \bibinfo{pages}{1–19}.
\newblock
\showISBNx{978-1-4503-8096-6}
\urldef\tempurl%
\url{https://doi.org/10.1145/3411764.3445188}
\showURL{%
\tempurl}


\bibitem[Ferrod et~al\mbox{.}(2021)]%
        {ferrod2021identifying}
\bibfield{author}{\bibinfo{person}{Roger Ferrod}, \bibinfo{person}{Federica
  Cena}, \bibinfo{person}{Luigi Di~Caro}, \bibinfo{person}{Dario Mana}, {and}
  \bibinfo{person}{Rossana~Grazia Simeoni}.} \bibinfo{year}{2021}\natexlab{}.
\newblock \showarticletitle{Identifying users' domain expertise from
  dialogues}. In \bibinfo{booktitle}{\emph{Adjunct {Proceedings} of the 29th
  {ACM} {Conference} on {User} {Modeling}, {Adaptation} and {Personalization}}}
  \emph{(\bibinfo{series}{{UMAP} '21})}. \bibinfo{publisher}{Association for
  Computing Machinery}, \bibinfo{address}{New York, NY, USA},
  \bibinfo{pages}{29--34}.
\newblock
\showISBNx{978-1-4503-8367-7}
\urldef\tempurl%
\url{https://doi.org/10.1145/3450614.3461683}
\showDOI{\tempurl}


\bibitem[Garland and Heckbert(1998)]%
        {garland1998quadrics}
\bibfield{author}{\bibinfo{person}{Michael Garland} {and}
  \bibinfo{person}{Paul~S. Heckbert}.} \bibinfo{year}{1998}\natexlab{}.
\newblock \showarticletitle{Simplifying surfaces with color and texture using
  quadric error metrics}. In \bibinfo{booktitle}{\emph{Proceedings of the
  conference on {Visualization} '98}} \emph{(\bibinfo{series}{{VIS} '98})}.
  \bibinfo{publisher}{IEEE Computer Society Press},
  \bibinfo{address}{Washington, DC, USA}, \bibinfo{pages}{263--269}.
\newblock
\showISBNx{978-1-58113-106-2}
\urldef\tempurl%
\url{https://doi.org/10.5555/288216.288280}
\showURL{%
\tempurl}


\bibitem[Garrett et~al\mbox{.}(2009)]%
        {garrett2009dim6}
\bibfield{author}{\bibinfo{person}{S.K. Garrett}, \bibinfo{person}{B.S.
  Caldwell}, \bibinfo{person}{E.C. Harris}, {and} \bibinfo{person}{M.C.
  Gonzalez}.} \bibinfo{year}{2009}\natexlab{}.
\newblock \showarticletitle{Six dimensions of expertise: a more comprehensive
  definition of cognitive expertise for team coordination}.
\newblock \bibinfo{journal}{\emph{Theoretical Issues in Ergonomics Science}}
  \bibinfo{volume}{10}, \bibinfo{number}{2} (\bibinfo{date}{March}
  \bibinfo{year}{2009}), \bibinfo{pages}{93--105}.
\newblock
\showISSN{1463-922X}
\urldef\tempurl%
\url{https://doi.org/10.1080/14639220802059190}
\showDOI{\tempurl}


\bibitem[Gigerenzer and Brighton(2009)]%
        {gigerenzer2009homo}
\bibfield{author}{\bibinfo{person}{Gerd Gigerenzer} {and}
  \bibinfo{person}{Henry Brighton}.} \bibinfo{year}{2009}\natexlab{}.
\newblock \showarticletitle{Homo heuristicus: why biased minds make better
  inferences}.
\newblock \bibinfo{journal}{\emph{Topics in Cognitive Science}}
  \bibinfo{volume}{1}, \bibinfo{number}{1} (\bibinfo{date}{Jan}
  \bibinfo{year}{2009}), \bibinfo{pages}{107–143}.
\newblock
\showISSN{1756-8765}
\urldef\tempurl%
\url{https://doi.org/10.1111/j.1756-8765.2008.01006.x}
\showDOI{\tempurl}


\bibitem[Gonz\'{a}lez et~al\mbox{.}(2017)]%
        {gonzalez2017pbo}
\bibfield{author}{\bibinfo{person}{Javier Gonz\'{a}lez},
  \bibinfo{person}{Zhenwen Dai}, \bibinfo{person}{Andreas Damianou}, {and}
  \bibinfo{person}{Neil~D. Lawrence}.} \bibinfo{year}{2017}\natexlab{}.
\newblock \showarticletitle{Preferential Bayesian Optimization}. In
  \bibinfo{booktitle}{\emph{Proceedings of the 34th International Conference on
  Machine Learning - Volume 70}} (Sydney, NSW, Australia)
  \emph{(\bibinfo{series}{ICML'17})}. \bibinfo{publisher}{JMLR.org},
  \bibinfo{address}{Australia}, \bibinfo{pages}{1282–1291}.
\newblock
\urldef\tempurl%
\url{https://doi.org/10.5555/3305381.3305514}
\showURL{%
\tempurl}


\bibitem[Grove and Meehl(1996)]%
        {grove1996comparative}
\bibfield{author}{\bibinfo{person}{William~M. Grove} {and}
  \bibinfo{person}{Paul~E. Meehl}.} \bibinfo{year}{1996}\natexlab{}.
\newblock \showarticletitle{Comparative efficiency of informal (subjective,
  impressionistic) and formal (mechanical, algorithmic) prediction procedures:
  The clinical–statistical controversy}.
\newblock \bibinfo{journal}{\emph{Psychology, Public Policy, and Law}}
  \bibinfo{volume}{2}, \bibinfo{number}{2} (\bibinfo{year}{1996}),
  \bibinfo{pages}{293–323}.
\newblock
\showISSN{1939-1528}
\urldef\tempurl%
\url{https://doi.org/10.1037/1076-8971.2.2.293}
\showDOI{\tempurl}


\bibitem[Grüne(2004)]%
        {gruene2004revealed}
\bibfield{author}{\bibinfo{person}{Till Grüne}.}
  \bibinfo{year}{2004}\natexlab{}.
\newblock \showarticletitle{The Problems of Testing Preference Axioms with
  Revealed Preference Theory}.
\newblock \bibinfo{journal}{\emph{Analyse \& Kritik}} \bibinfo{volume}{26},
  \bibinfo{number}{2} (\bibinfo{date}{Nov} \bibinfo{year}{2004}),
  \bibinfo{pages}{382–397}.
\newblock
\showISSN{2365-9858}
\urldef\tempurl%
\url{https://doi.org/10.1515/auk-2004-0204}
\showDOI{\tempurl}


\bibitem[Hausman(2011)]%
        {hausman2011preference}
\bibfield{author}{\bibinfo{person}{Daniel~M. Hausman}.}
  \bibinfo{year}{2011}\natexlab{}.
\newblock \bibinfo{booktitle}{\emph{Preference, Value, Choice, and Welfare}}.
\newblock \bibinfo{publisher}{Cambridge University Press},
  \bibinfo{address}{New York, USA}.
\newblock
\showISBNx{978-1-139-50537-6}


\bibitem[Holliday et~al\mbox{.}(2016)]%
        {holliday2016time}
\bibfield{author}{\bibinfo{person}{Daniel Holliday}, \bibinfo{person}{Stephanie
  Wilson}, {and} \bibinfo{person}{Simone Stumpf}.}
  \bibinfo{year}{2016}\natexlab{}.
\newblock \showarticletitle{User Trust in Intelligent Systems: A Journey Over
  Time}. In \bibinfo{booktitle}{\emph{Proceedings of the 21st International
  Conference on Intelligent User Interfaces}} (Sonoma, California, USA)
  \emph{(\bibinfo{series}{IUI '16})}. \bibinfo{publisher}{Association for
  Computing Machinery}, \bibinfo{address}{New York, NY, USA},
  \bibinfo{pages}{164–168}.
\newblock
\showISBNx{9781450341370}
\urldef\tempurl%
\url{https://doi.org/10.1145/2856767.2856811}
\showDOI{\tempurl}


\bibitem[Holtzman et~al\mbox{.}(2019)]%
        {holtzman2020topp}
\bibfield{author}{\bibinfo{person}{Ari Holtzman}, \bibinfo{person}{Jan Buys},
  \bibinfo{person}{Li Du}, \bibinfo{person}{Maxwell Forbes}, {and}
  \bibinfo{person}{Yejin Choi}.} \bibinfo{year}{2019}\natexlab{}.
\newblock \bibinfo{title}{The Curious Case of Neural Text Degeneration}.
\newblock
\newblock
\urldef\tempurl%
\url{https://doi.org/10.48550/ARXIV.1904.09751}
\showDOI{\tempurl}


\bibitem[Iyengar et~al\mbox{.}(2006)]%
        {iyengar2006doing}
\bibfield{author}{\bibinfo{person}{Sheena~S Iyengar},
  \bibinfo{person}{Rachael~E Wells}, {and} \bibinfo{person}{Barry Schwartz}.}
  \bibinfo{year}{2006}\natexlab{}.
\newblock \showarticletitle{Doing better but feeling worse: Looking for the
  “best” job undermines satisfaction}.
\newblock \bibinfo{journal}{\emph{Psychological Science}} \bibinfo{volume}{17},
  \bibinfo{number}{2} (\bibinfo{year}{2006}), \bibinfo{pages}{143--150}.
\newblock
\urldef\tempurl%
\url{https://doi.org/10.1111/j.1467-9280.2006.01677.x}
\showDOI{\tempurl}


\bibitem[Jakob et~al\mbox{.}(2015)]%
        {jakob2015instant}
\bibfield{author}{\bibinfo{person}{Wenzel Jakob}, \bibinfo{person}{Marco
  Tarini}, \bibinfo{person}{Daniele Panozzo}, {and} \bibinfo{person}{Olga
  Sorkine-Hornung}.} \bibinfo{year}{2015}\natexlab{}.
\newblock \showarticletitle{Instant {Field}-{Aligned} {Meshes}}.
\newblock \bibinfo{journal}{\emph{ACM Transactions on Graphics}}
  \bibinfo{volume}{34}, \bibinfo{number}{6} (\bibinfo{date}{Oct.}
  \bibinfo{year}{2015}), \bibinfo{pages}{189:1--189:15}.
\newblock
\showISSN{0730-0301}
\urldef\tempurl%
\url{https://doi.org/10.1145/2816795.2818078}
\showURL{%
\tempurl}


\bibitem[Jannach et~al\mbox{.}(2010)]%
        {jannach2010recommender}
\bibfield{author}{\bibinfo{person}{Dietmar Jannach}, \bibinfo{person}{Markus
  Zanker}, \bibinfo{person}{Alexander Felfernig}, {and}
  \bibinfo{person}{Gerhard Friedrich}.} \bibinfo{year}{2010}\natexlab{}.
\newblock \bibinfo{booktitle}{\emph{Recommender Systems: An Introduction}}.
\newblock \bibinfo{publisher}{Cambridge University Press},
  \bibinfo{address}{New York, USA}.
\newblock
\showISBNx{978-1-139-49259-1}


\bibitem[Jerry~Lin et~al\mbox{.}(2022)]%
        {lin2022preference}
\bibfield{author}{\bibinfo{person}{Zhiyuan Jerry~Lin}, \bibinfo{person}{Raul
  Astudillo}, \bibinfo{person}{Peter Frazier}, {and} \bibinfo{person}{Eytan
  Bakshy}.} \bibinfo{year}{2022}\natexlab{}.
\newblock \showarticletitle{Preference Exploration for Efficient Bayesian
  Optimization with Multiple Outcomes}. In
  \bibinfo{booktitle}{\emph{Proceedings of The 25th International Conference on
  Artificial Intelligence and Statistics}} \emph{(\bibinfo{series}{Proceedings
  of Machine Learning Research}, Vol.~\bibinfo{volume}{151})},
  \bibfield{editor}{\bibinfo{person}{Gustau Camps-Valls},
  \bibinfo{person}{Francisco J.~R. Ruiz}, {and} \bibinfo{person}{Isabel
  Valera}} (Eds.). \bibinfo{publisher}{PMLR}, \bibinfo{address}{Virtual},
  \bibinfo{pages}{4235--4258}.
\newblock
\urldef\tempurl%
\url{https://proceedings.mlr.press/v151/jerry-lin22a.html}
\showURL{%
\tempurl}


\bibitem[Kahneman et~al\mbox{.}(2021)]%
        {kahneman2021noise}
\bibfield{author}{\bibinfo{person}{Daniel Kahneman}, \bibinfo{person}{Olivier
  Sibony}, {and} \bibinfo{person}{Cass~R. Sunstein}.}
  \bibinfo{year}{2021}\natexlab{}.
\newblock \bibinfo{booktitle}{\emph{Noise}}.
\newblock \bibinfo{publisher}{HarperCollins UK}, \bibinfo{address}{UK}.
\newblock
\showISBNx{978-0-00-830901-5}


\bibitem[Kizilcec(2016)]%
        {kizilcec2016trust}
\bibfield{author}{\bibinfo{person}{René~F. Kizilcec}.}
  \bibinfo{year}{2016}\natexlab{}.
\newblock \showarticletitle{How Much Information? Effects of Transparency on
  Trust in an Algorithmic Interface}. In \bibinfo{booktitle}{\emph{Proceedings
  of the 2016 CHI Conference on Human Factors in Computing Systems}}
  \emph{(\bibinfo{series}{CHI ’16})}. \bibinfo{publisher}{Association for
  Computing Machinery}, \bibinfo{address}{New York, NY, USA},
  \bibinfo{pages}{2390–2395}.
\newblock
\showISBNx{978-1-4503-3362-7}
\urldef\tempurl%
\url{https://doi.org/10.1145/2858036.2858402}
\showDOI{\tempurl}


\bibitem[Knupp(2000)]%
        {knupp2000fem}
\bibfield{author}{\bibinfo{person}{Patrick~M. Knupp}.}
  \bibinfo{year}{2000}\natexlab{}.
\newblock \showarticletitle{Achieving finite element mesh quality via
  optimization of the Jacobian matrix norm and associated quantities. Part
  I—a framework for surface mesh optimization}.
\newblock \bibinfo{journal}{\emph{Internat. J. Numer. Methods Engrg.}}
  \bibinfo{volume}{48}, \bibinfo{number}{3} (\bibinfo{year}{2000}),
  \bibinfo{pages}{401–420}.
\newblock
\showISSN{1097-0207}
\urldef\tempurl%
\url{https://doi.org/10.1002/(SICI)1097-0207(20000530)48:3<401::AID-NME880>3.0.CO;2-D}
\showDOI{\tempurl}


\bibitem[Korhonen and You(2012)]%
        {korhonen2012peak}
\bibfield{author}{\bibinfo{person}{Jari Korhonen} {and}
  \bibinfo{person}{Junyong You}.} \bibinfo{year}{2012}\natexlab{}.
\newblock \showarticletitle{Peak signal-to-noise ratio revisited: Is simple
  beautiful?}. In \bibinfo{booktitle}{\emph{2012 Fourth International Workshop
  on Quality of Multimedia Experience}}. \bibinfo{publisher}{IEEE},
  \bibinfo{address}{Melbourne, VIC, Australia}, \bibinfo{pages}{37–38}.
\newblock
\urldef\tempurl%
\url{https://doi.org/10.1109/QoMEX.2012.6263880}
\showDOI{\tempurl}


\bibitem[Kotzee and Smit(2018)]%
        {kotzee2018dim2}
\bibfield{author}{\bibinfo{person}{Ben Kotzee} {and} \bibinfo{person}{Jp
  Smit}.} \bibinfo{year}{2018}\natexlab{}.
\newblock \showarticletitle{Two Social Dimensions of Expertise}.
\newblock In \bibinfo{booktitle}{\emph{Education and Expertise}}.
  \bibinfo{publisher}{John Wiley \& Sons, Ltd}, \bibinfo{address}{New Jersey,
  USA}, \bibinfo{pages}{99–116}.
\newblock
\showISBNx{978-1-119-52726-8}
\urldef\tempurl%
\url{https://doi.org/10.1002/9781119527268.ch5}
\showDOI{\tempurl}


\bibitem[Koyama et~al\mbox{.}(2016)]%
        {koyama2016selph}
\bibfield{author}{\bibinfo{person}{Yuki Koyama}, \bibinfo{person}{Daisuke
  Sakamoto}, {and} \bibinfo{person}{Takeo Igarashi}.}
  \bibinfo{year}{2016}\natexlab{}.
\newblock \showarticletitle{SelPh: Progressive Learning and Support of Manual
  Photo Color Enhancement}. In \bibinfo{booktitle}{\emph{Proceedings of the
  2016 CHI Conference on Human Factors in Computing Systems}} (San Jose,
  California, USA) \emph{(\bibinfo{series}{CHI '16})}.
  \bibinfo{publisher}{Association for Computing Machinery},
  \bibinfo{address}{New York, NY, USA}, \bibinfo{pages}{2520–2532}.
\newblock
\showISBNx{9781450333627}
\urldef\tempurl%
\url{https://doi.org/10.1145/2858036.2858111}
\showDOI{\tempurl}


\bibitem[Koyama et~al\mbox{.}(2020)]%
        {koyam2020seqgallery}
\bibfield{author}{\bibinfo{person}{Yuki Koyama}, \bibinfo{person}{Issei Sato},
  {and} \bibinfo{person}{Masataka Goto}.} \bibinfo{year}{2020}\natexlab{}.
\newblock \showarticletitle{Sequential gallery for interactive visual design
  optimization}.
\newblock \bibinfo{journal}{\emph{ACM Transactions on Graphics}}
  \bibinfo{volume}{39}, \bibinfo{number}{4} (\bibinfo{date}{July}
  \bibinfo{year}{2020}), \bibinfo{pages}{88:88:1--88:88:12}.
\newblock
\showISSN{0730-0301}
\urldef\tempurl%
\url{https://doi.org/10.1145/3386569.3392444}
\showURL{%
\tempurl}


\bibitem[Koyama et~al\mbox{.}(2017)]%
        {koyama2017sequential}
\bibfield{author}{\bibinfo{person}{Yuki Koyama}, \bibinfo{person}{Issei Sato},
  \bibinfo{person}{Daisuke Sakamoto}, {and} \bibinfo{person}{Takeo Igarashi}.}
  \bibinfo{year}{2017}\natexlab{}.
\newblock \showarticletitle{Sequential Line Search for Efficient Visual Design
  Optimization by Crowds}.
\newblock \bibinfo{journal}{\emph{ACM Trans. Graph.}} \bibinfo{volume}{36},
  \bibinfo{number}{4}, Article \bibinfo{articleno}{48} (\bibinfo{date}{jul}
  \bibinfo{year}{2017}), \bibinfo{numpages}{11}~pages.
\newblock
\showISSN{0730-0301}
\urldef\tempurl%
\url{https://doi.org/10.1145/3072959.3073598}
\showDOI{\tempurl}


\bibitem[Krishna and Morgan(2001)]%
        {krishna2001model}
\bibfield{author}{\bibinfo{person}{Vijay Krishna} {and} \bibinfo{person}{John
  Morgan}.} \bibinfo{year}{2001}\natexlab{}.
\newblock \showarticletitle{A {Model} of {Expertise}}.
\newblock \bibinfo{journal}{\emph{The Quarterly Journal of Economics}}
  \bibinfo{volume}{116}, \bibinfo{number}{2} (\bibinfo{date}{May}
  \bibinfo{year}{2001}), \bibinfo{pages}{747--775}.
\newblock
\showISSN{0033-5533}
\urldef\tempurl%
\url{https://doi.org/10.1162/00335530151144159}
\showDOI{\tempurl}


\bibitem[Kuznetsova et~al\mbox{.}(2017)]%
        {kuznetsova2017lmertest}
\bibfield{author}{\bibinfo{person}{Alexandra Kuznetsova},
  \bibinfo{person}{Per~B. Brockhoff}, {and} \bibinfo{person}{Rune H.~B.
  Christensen}.} \bibinfo{year}{2017}\natexlab{}.
\newblock \showarticletitle{{lmerTest} Package: Tests in Linear Mixed Effects
  Models}.
\newblock \bibinfo{journal}{\emph{Journal of Statistical Software}}
  \bibinfo{volume}{82}, \bibinfo{number}{13} (\bibinfo{year}{2017}),
  \bibinfo{pages}{1--26}.
\newblock
\urldef\tempurl%
\url{https://doi.org/10.18637/jss.v082.i13}
\showDOI{\tempurl}


\bibitem[Lee et~al\mbox{.}(2012)]%
        {lee2012ranking}
\bibfield{author}{\bibinfo{person}{Michael~D. Lee}, \bibinfo{person}{Mark
  Steyvers}, \bibinfo{person}{Mindy de Young}, {and} \bibinfo{person}{Brent
  Miller}.} \bibinfo{year}{2012}\natexlab{}.
\newblock \showarticletitle{Inferring Expertise in Knowledge and Prediction
  Ranking Tasks}.
\newblock \bibinfo{journal}{\emph{Topics in Cognitive Science}}
  \bibinfo{volume}{4}, \bibinfo{number}{1} (\bibinfo{year}{2012}),
  \bibinfo{pages}{151–163}.
\newblock
\showISSN{1756-8765}
\urldef\tempurl%
\url{https://doi.org/10.1111/j.1756-8765.2011.01175.x}
\showDOI{\tempurl}


\bibitem[Lescoat et~al\mbox{.}(2020)]%
        {lescoat2020specmeshsim}
\bibfield{author}{\bibinfo{person}{Thibault Lescoat},
  \bibinfo{person}{Hsueh-Ti~Derek Liu}, \bibinfo{person}{Jean-Marc Thiery},
  \bibinfo{person}{Alec Jacobson}, \bibinfo{person}{Tamy Boubekeur}, {and}
  \bibinfo{person}{Maks Ovsjanikov}.} \bibinfo{year}{2020}\natexlab{}.
\newblock \showarticletitle{Spectral Mesh Simplification}.
\newblock \bibinfo{journal}{\emph{Computer Graphics Forum}}
  \bibinfo{volume}{39}, \bibinfo{number}{2} (\bibinfo{year}{2020}),
  \bibinfo{pages}{315–324}.
\newblock
\showISSN{1467-8659}
\urldef\tempurl%
\url{https://doi.org/10.1111/cgf.13932}
\showDOI{\tempurl}


\bibitem[Letham et~al\mbox{.}(2019)]%
        {letham2019constrained}
\bibfield{author}{\bibinfo{person}{Benjamin Letham}, \bibinfo{person}{Brian
  Karrer}, \bibinfo{person}{Guilherme Ottoni}, {and} \bibinfo{person}{Eytan
  Bakshy}.} \bibinfo{year}{2019}\natexlab{}.
\newblock \showarticletitle{Constrained Bayesian Optimization with Noisy
  Experiments}.
\newblock \bibinfo{journal}{\emph{Bayesian Analysis}} \bibinfo{volume}{14},
  \bibinfo{number}{2} (\bibinfo{date}{Jun} \bibinfo{year}{2019}),
  \bibinfo{pages}{495–519}.
\newblock
\showISSN{1936-0975, 1931-6690}
\urldef\tempurl%
\url{https://doi.org/10.1214/18-BA1110}
\showDOI{\tempurl}


\bibitem[Lin(2004)]%
        {lin2004rouge}
\bibfield{author}{\bibinfo{person}{Chin-Yew Lin}.}
  \bibinfo{year}{2004}\natexlab{}.
\newblock \showarticletitle{ROUGE: A Package for Automatic Evaluation of
  Summaries}. In \bibinfo{booktitle}{\emph{Text Summarization Branches Out}}.
  \bibinfo{publisher}{Association for Computational Linguistics},
  \bibinfo{address}{Barcelona, Spain}, \bibinfo{pages}{74–81}.
\newblock
\urldef\tempurl%
\url{https://aclanthology.org/W04-1013}
\showURL{%
\tempurl}


\bibitem[Marks et~al\mbox{.}(1997)]%
        {marks1997galleries}
\bibfield{author}{\bibinfo{person}{J. Marks}, \bibinfo{person}{B. Andalman},
  \bibinfo{person}{P.~A. Beardsley}, \bibinfo{person}{W. Freeman},
  \bibinfo{person}{S. Gibson}, \bibinfo{person}{J. Hodgins},
  \bibinfo{person}{T. Kang}, \bibinfo{person}{B. Mirtich}, \bibinfo{person}{H.
  Pfister}, \bibinfo{person}{W. Ruml}, \bibinfo{person}{K. Ryall},
  \bibinfo{person}{J. Seims}, {and} \bibinfo{person}{S. Shieber}.}
  \bibinfo{year}{1997}\natexlab{}.
\newblock \showarticletitle{{Design Galleries: A General Approach to Setting
  Parameters for Computer Graphics and Animation}}. In
  \bibinfo{booktitle}{\emph{Proceedings of the 24th annual conference on
  {Computer} graphics and interactive techniques}}
  \emph{(\bibinfo{series}{{SIGGRAPH} '97})}. \bibinfo{publisher}{ACM
  Press/Addison-Wesley Publishing Co.}, \bibinfo{address}{USA},
  \bibinfo{pages}{389--400}.
\newblock
\showISBNx{978-0-89791-896-1}
\urldef\tempurl%
\url{https://doi.org/10.1145/258734.258887}
\showURL{%
\tempurl}


\bibitem[Mikkola et~al\mbox{.}(2020)]%
        {mikkola2020projective}
\bibfield{author}{\bibinfo{person}{Petrus Mikkola}, \bibinfo{person}{Milica
  Todorovi{\'c}}, \bibinfo{person}{Jari J{\"a}rvi}, \bibinfo{person}{Patrick
  Rinke}, {and} \bibinfo{person}{Samuel Kaski}.}
  \bibinfo{year}{2020}\natexlab{}.
\newblock \showarticletitle{Projective Preferential Bayesian Optimization}. In
  \bibinfo{booktitle}{\emph{International Conference on Machine Learning}}.
  PMLR, \bibinfo{publisher}{MLResearchPress}, \bibinfo{address}{Online},
  \bibinfo{pages}{6884--6892}.
\newblock
\urldef\tempurl%
\url{https://doi.org/10.5555/3524938.3525577}
\showDOI{\tempurl}


\bibitem[Miller(2019)]%
        {miller2019creative}
\bibfield{author}{\bibinfo{person}{Arthur~I. Miller}.}
  \bibinfo{year}{2019}\natexlab{}.
\newblock \bibinfo{booktitle}{\emph{The Artist in the Machine: The World of
  AI-Powered Creativity}}.
\newblock \bibinfo{publisher}{MIT Press}, \bibinfo{address}{Cambridge,
  Massachusetts, USA}.
\newblock
\showISBNx{978-0-262-35460-8}


\bibitem[Monarch(2021)]%
        {monarch2021hitl}
\bibfield{author}{\bibinfo{person}{Robert~M. Monarch}.}
  \bibinfo{year}{2021}\natexlab{}.
\newblock \bibinfo{booktitle}{\emph{Human-in-the-Loop Machine Learning: Active
  Learning and Annotation for Human-centered AI}}.
\newblock \bibinfo{publisher}{Simon and Schuster}, \bibinfo{address}{USA}.
\newblock
\showISBNx{978-1-61729-674-1}


\bibitem[Ooge and Verbert(2021)]%
        {ooge2021trust}
\bibfield{author}{\bibinfo{person}{Jeroen Ooge} {and} \bibinfo{person}{Katrien
  Verbert}.} \bibinfo{year}{2021}\natexlab{}.
\newblock \bibinfo{title}{Trust in {Prediction} {Models}: a {Mixed}-{Methods}
  {Pilot} {Study} on the {Impact} of {Domain} {Expertise}}.
\newblock
\newblock
\urldef\tempurl%
\url{https://doi.org/10.48550/arXiv.2109.08183}
\showDOI{\tempurl}
\showeprint{arXiv:2109.08183}


\bibitem[Ou et~al\mbox{.}(2022)]%
        {ou2022infloop}
\bibfield{author}{\bibinfo{person}{Changkun Ou}, \bibinfo{person}{Daniel
  Buschek}, \bibinfo{person}{Sven Mayer}, {and} \bibinfo{person}{Andreas
  Butz}.} \bibinfo{year}{2022}\natexlab{}.
\newblock \showarticletitle{The Human in the Infinite Loop: A Case Study on
  Revealing and Explaining Human-AI Interaction Loop Failures}. In
  \bibinfo{booktitle}{\emph{Proceedings of Mensch Und Computer 2022}}
  \emph{(\bibinfo{series}{MuC '22})}. \bibinfo{publisher}{Association for
  Computing Machinery}, \bibinfo{address}{Darmstadt, Germany},
  \bibinfo{pages}{1--11}.
\newblock
\urldef\tempurl%
\url{https://doi.org/10.1145/3543758.3543761}
\showDOI{\tempurl}


\bibitem[Papineni et~al\mbox{.}(2002)]%
        {papineni2002bleu}
\bibfield{author}{\bibinfo{person}{Kishore Papineni}, \bibinfo{person}{Salim
  Roukos}, \bibinfo{person}{Todd Ward}, {and} \bibinfo{person}{Wei-Jing Zhu}.}
  \bibinfo{year}{2002}\natexlab{}.
\newblock \showarticletitle{Bleu: a Method for Automatic Evaluation of Machine
  Translation}. In \bibinfo{booktitle}{\emph{Proceedings of the 40th Annual
  Meeting of the Association for Computational Linguistics}}.
  \bibinfo{publisher}{Association for Computational Linguistics},
  \bibinfo{address}{Philadelphia, Pennsylvania, USA},
  \bibinfo{pages}{311–318}.
\newblock
\urldef\tempurl%
\url{https://doi.org/10.3115/1073083.1073135}
\showDOI{\tempurl}


\bibitem[Pareto(1912)]%
        {pareto1912manuel}
\bibfield{author}{\bibinfo{person}{Vilfredo Pareto}.}
  \bibinfo{year}{1912}\natexlab{}.
\newblock \showarticletitle{Manuel d’{\'e}conomie politique}.
\newblock \bibinfo{journal}{\emph{Bull. Amer. Math. Soc}} \bibinfo{volume}{18},
  \bibinfo{number}{462-474} (\bibinfo{year}{1912}), \bibinfo{pages}{3}.
\newblock


\bibitem[Rezende et~al\mbox{.}(2014)]%
        {rezende2014stochastic}
\bibfield{author}{\bibinfo{person}{Danilo~Jimenez Rezende},
  \bibinfo{person}{Shakir Mohamed}, {and} \bibinfo{person}{Daan Wierstra}.}
  \bibinfo{year}{2014}\natexlab{}.
\newblock \showarticletitle{Stochastic Backpropagation and Approximate
  Inference in Deep Generative Models}. In
  \bibinfo{booktitle}{\emph{Proceedings of the 31st International Conference on
  International Conference on Machine Learning - Volume 32}} (Beijing, China)
  \emph{(\bibinfo{series}{ICML'14})}. \bibinfo{publisher}{JMLR.org},
  \bibinfo{address}{Beijing, China}, \bibinfo{pages}{II–1278–II–1286}.
\newblock
\urldef\tempurl%
\url{https://proceedings.mlr.press/v32/rezende14.html}
\showURL{%
\tempurl}


\bibitem[Rooderkerk et~al\mbox{.}(2011)]%
        {rooderkerk2011incorporating}
\bibfield{author}{\bibinfo{person}{Robert~P. Rooderkerk},
  \bibinfo{person}{Harald~J. Van~Heerde}, {and} \bibinfo{person}{Tammo~H.A.
  Bijmolt}.} \bibinfo{year}{2011}\natexlab{}.
\newblock \showarticletitle{Incorporating Context Effects into a Choice Model}.
\newblock \bibinfo{journal}{\emph{Journal of Marketing Research}}
  \bibinfo{volume}{48}, \bibinfo{number}{4} (\bibinfo{date}{Aug}
  \bibinfo{year}{2011}), \bibinfo{pages}{767–780}.
\newblock
\showISSN{0022-2437}
\urldef\tempurl%
\url{https://doi.org/10.1509/jmkr.48.4.767}
\showDOI{\tempurl}


\bibitem[Schmidtz(2004)]%
        {schmidtz2004satisficing}
\bibfield{author}{\bibinfo{person}{David Schmidtz}.}
  \bibinfo{year}{2004}\natexlab{}.
\newblock \showarticletitle{Satisficing as a Humanly Rational Strategy}.
\newblock In \bibinfo{booktitle}{\emph{Satisficing and Maximizing: Moral
  Theorists on Practical Reason}}, \bibfield{editor}{\bibinfo{person}{Michael
  Byron}} (Ed.). \bibinfo{publisher}{Cambridge University Press},
  \bibinfo{address}{New York, USA}, \bibinfo{pages}{30–59}.
\newblock
\urldef\tempurl%
\url{https://doi.org/10.1017/CBO9780511617058.003}
\showDOI{\tempurl}


\bibitem[Schulz et~al\mbox{.}(2018)]%
        {schulz2018interactive}
\bibfield{author}{\bibinfo{person}{Adriana Schulz}, \bibinfo{person}{Harrison
  Wang}, \bibinfo{person}{Eitan Grinspun}, \bibinfo{person}{Justin Solomon},
  {and} \bibinfo{person}{Wojciech Matusik}.} \bibinfo{year}{2018}\natexlab{}.
\newblock \showarticletitle{Interactive exploration of design trade-offs}.
\newblock \bibinfo{journal}{\emph{ACM Transactions on Graphics}}
  \bibinfo{volume}{37}, \bibinfo{number}{4} (\bibinfo{date}{Jul}
  \bibinfo{year}{2018}), \bibinfo{pages}{131:1--131:14}.
\newblock
\showISSN{0730-0301}
\urldef\tempurl%
\url{https://doi.org/10.1145/3197517.3201385}
\showDOI{\tempurl}


\bibitem[Schwartz et~al\mbox{.}(2002)]%
        {schwartz2002maximizing}
\bibfield{author}{\bibinfo{person}{Barry Schwartz}, \bibinfo{person}{Andrew
  Ward}, \bibinfo{person}{John Monterosso}, \bibinfo{person}{Sonja
  Lyubomirsky}, \bibinfo{person}{Katherine White}, {and}
  \bibinfo{person}{Darrin~R. Lehman}.} \bibinfo{year}{2002}\natexlab{}.
\newblock \showarticletitle{Maximizing versus satisficing: Happiness is a
  matter of choice.}
\newblock \bibinfo{journal}{\emph{Journal of Personality and Social
  Psychology}} \bibinfo{volume}{83}, \bibinfo{number}{5} (\bibinfo{date}{Nov}
  \bibinfo{year}{2002}), \bibinfo{pages}{1178–1197}.
\newblock
\showISSN{1939-1315, 0022-3514}
\urldef\tempurl%
\url{https://doi.org/10.1037/0022-3514.83.5.1178}
\showDOI{\tempurl}


\bibitem[Shahriari et~al\mbox{.}(2016)]%
        {shahriari2016boreview}
\bibfield{author}{\bibinfo{person}{Bobak Shahriari}, \bibinfo{person}{Kevin
  Swersky}, \bibinfo{person}{Ziyu Wang}, \bibinfo{person}{Ryan~P. Adams}, {and}
  \bibinfo{person}{Nando de Freitas}.} \bibinfo{year}{2016}\natexlab{}.
\newblock \showarticletitle{Taking the {Human} {Out} of the {Loop}: {A}
  {Review} of {Bayesian} {Optimization}}.
\newblock \bibinfo{journal}{\emph{Proc. IEEE}} \bibinfo{volume}{104},
  \bibinfo{number}{1} (\bibinfo{date}{Jan.} \bibinfo{year}{2016}),
  \bibinfo{pages}{148--175}.
\newblock
\showISSN{1558-2256}
\urldef\tempurl%
\url{https://doi.org/10.1109/JPROC.2015.2494218}
\showDOI{\tempurl}


\bibitem[Siivola et~al\mbox{.}(2021)]%
        {siivola2021pbbo}
\bibfield{author}{\bibinfo{person}{Eero Siivola}, \bibinfo{person}{Akash~Kumar
  Dhaka}, \bibinfo{person}{Michael~Riis Andersen}, \bibinfo{person}{Javier
  González}, \bibinfo{person}{Pablo~García Moreno}, {and}
  \bibinfo{person}{Aki Vehtari}.} \bibinfo{year}{2021}\natexlab{}.
\newblock \showarticletitle{Preferential Batch Bayesian Optimization}. In
  \bibinfo{booktitle}{\emph{2021 IEEE 31st International Workshop on Machine
  Learning for Signal Processing (MLSP)}}. \bibinfo{publisher}{IEEE},
  \bibinfo{address}{Gold Coast, Australia}, \bibinfo{pages}{1–6}.
\newblock
\showISSN{1551-2541}
\urldef\tempurl%
\url{https://doi.org/10.1109/MLSP52302.2021.9596494}
\showDOI{\tempurl}


\bibitem[Simon(1955)]%
        {simon1955rational}
\bibfield{author}{\bibinfo{person}{Herbert~A. Simon}.}
  \bibinfo{year}{1955}\natexlab{}.
\newblock \showarticletitle{A {Behavioral} {Model} of {Rational} {Choice}}.
\newblock \bibinfo{journal}{\emph{The Quarterly Journal of Economics}}
  \bibinfo{volume}{69}, \bibinfo{number}{1} (\bibinfo{date}{Feb.}
  \bibinfo{year}{1955}), \bibinfo{pages}{99–118}.
\newblock
\showISSN{0033-5533}
\urldef\tempurl%
\url{https://doi.org/10.2307/1884852}
\showURL{%
\tempurl}


\bibitem[Simpson et~al\mbox{.}(2020)]%
        {simpson2020interactive}
\bibfield{author}{\bibinfo{person}{Edwin Simpson}, \bibinfo{person}{Yang Gao},
  {and} \bibinfo{person}{Iryna Gurevych}.} \bibinfo{year}{2020}\natexlab{}.
\newblock \showarticletitle{{Interactive Text Ranking with Bayesian
  Optimization: A Case Study on Community QA and Summarization}}.
\newblock \bibinfo{journal}{\emph{Transactions of the Association for
  Computational Linguistics}}  \bibinfo{volume}{8} (\bibinfo{date}{12}
  \bibinfo{year}{2020}), \bibinfo{pages}{759--775}.
\newblock
\showISSN{2307-387X}
\urldef\tempurl%
\url{https://doi.org/10.1162/tacl_a_00344}
\showDOI{\tempurl}


\bibitem[Thurstone(1927)]%
        {thurstone1927comparelaw}
\bibfield{author}{\bibinfo{person}{Louis~L. Thurstone}.}
  \bibinfo{year}{1927}\natexlab{}.
\newblock \showarticletitle{A law of comparative judgment}.
\newblock \bibinfo{journal}{\emph{Psychological Review}} \bibinfo{volume}{34},
  \bibinfo{number}{4} (\bibinfo{year}{1927}), \bibinfo{pages}{273--286}.
\newblock
\showISSN{1939-1471(Electronic),0033-295X(Print)}
\urldef\tempurl%
\url{https://doi.org/10.1037/h0070288}
\showURL{%
\tempurl}


\bibitem[Treem and Leonardi(2016)]%
        {treem2016communicate}
\bibfield{author}{\bibinfo{person}{Jeffrey~W. Treem} {and}
  \bibinfo{person}{Paul~M. Leonardi}.} \bibinfo{year}{2016}\natexlab{}.
\newblock \bibinfo{booktitle}{\emph{Expertise, {Communication}, and
  {Organizing}}}.
\newblock \bibinfo{publisher}{Oxford University Press}, \bibinfo{address}{New
  York, NY, USA}.
\newblock
\showISBNx{978-0-19-105974-2}


\bibitem[Tversky and Kahneman(1974)]%
        {kahneman1974judgment}
\bibfield{author}{\bibinfo{person}{Amos Tversky} {and} \bibinfo{person}{Daniel
  Kahneman}.} \bibinfo{year}{1974}\natexlab{}.
\newblock \showarticletitle{{Judgment under Uncertainty: Heuristics and
  Biases}}.
\newblock \bibinfo{journal}{\emph{Science}} \bibinfo{volume}{185},
  \bibinfo{number}{4157} (\bibinfo{date}{Sept.} \bibinfo{year}{1974}),
  \bibinfo{pages}{1124--1131}.
\newblock
\showISSN{0036-8075, 1095-9203}
\urldef\tempurl%
\url{https://doi.org/10.1126/science.185.4157.1124}
\showURL{%
\tempurl}


\bibitem[Wang et~al\mbox{.}(2004)]%
        {wang2004ssim}
\bibfield{author}{\bibinfo{person}{Zhou Wang}, \bibinfo{person}{A.C. Bovik},
  \bibinfo{person}{H.R. Sheikh}, {and} \bibinfo{person}{E.P. Simoncelli}.}
  \bibinfo{year}{2004}\natexlab{}.
\newblock \showarticletitle{Image quality assessment: from error visibility to
  structural similarity}.
\newblock \bibinfo{journal}{\emph{IEEE Transactions on Image Processing}}
  \bibinfo{volume}{13}, \bibinfo{number}{4} (\bibinfo{date}{Apr}
  \bibinfo{year}{2004}), \bibinfo{pages}{600–612}.
\newblock
\showISSN{1941-0042}
\urldef\tempurl%
\url{https://doi.org/10.1109/TIP.2003.819861}
\showDOI{\tempurl}


\bibitem[Wilson et~al\mbox{.}(2017)]%
        {wilson2017reparam}
\bibfield{author}{\bibinfo{person}{James~T. Wilson}, \bibinfo{person}{Riccardo
  Moriconi}, \bibinfo{person}{Frank Hutter}, {and} \bibinfo{person}{Marc~Peter
  Deisenroth}.} \bibinfo{year}{2017}\natexlab{}.
\newblock \bibinfo{title}{The reparameterization trick for acquisition
  functions}.
\newblock
\newblock
\urldef\tempurl%
\url{https://doi.org/10.48550/ARXIV.1712.00424}
\showDOI{\tempurl}


\bibitem[Wobbrock et~al\mbox{.}(2011)]%
        {wobbrock2011art}
\bibfield{author}{\bibinfo{person}{Jacob~O. Wobbrock}, \bibinfo{person}{Leah
  Findlater}, \bibinfo{person}{Darren Gergle}, {and} \bibinfo{person}{James~J.
  Higgins}.} \bibinfo{year}{2011}\natexlab{}.
\newblock \showarticletitle{The Aligned Rank Transform for Nonparametric
  Factorial Analyses Using Only Anova Procedures}. In
  \bibinfo{booktitle}{\emph{Proceedings of the SIGCHI Conference on Human
  Factors in Computing Systems}} (Vancouver, BC, Canada)
  \emph{(\bibinfo{series}{CHI '11})}. \bibinfo{publisher}{Association for
  Computing Machinery}, \bibinfo{address}{New York, NY, USA},
  \bibinfo{pages}{143–146}.
\newblock
\showISBNx{9781450302289}
\urldef\tempurl%
\url{https://doi.org/10.1145/1978942.1978963}
\showDOI{\tempurl}


\bibitem[Yue et~al\mbox{.}(2021)]%
        {koyama2021ui}
\bibfield{author}{\bibinfo{person}{Yonghao Yue}, \bibinfo{person}{Yuki Koyama},
  \bibinfo{person}{Issei Sato}, {and} \bibinfo{person}{Takeo Igarashi}.}
  \bibinfo{year}{2021}\natexlab{}.
\newblock \showarticletitle{User interfaces for high-dimensional design
  problems: from theories to implementations}. In \bibinfo{booktitle}{\emph{ACM
  SIGGRAPH 2021 Courses}} \emph{(\bibinfo{series}{SIGGRAPH ’21})}.
  \bibinfo{publisher}{Association for Computing Machinery},
  \bibinfo{address}{New York, NY, USA}, \bibinfo{pages}{1–34}.
\newblock
\showISBNx{978-1-4503-8361-5}
\urldef\tempurl%
\url{https://doi.org/10.1145/3450508.3464551}
\showDOI{\tempurl}


\bibitem[Zhou et~al\mbox{.}(2021)]%
        {yijun2021melody}
\bibfield{author}{\bibinfo{person}{Yijun Zhou}, \bibinfo{person}{Yuki Koyama},
  \bibinfo{person}{Masataka Goto}, {and} \bibinfo{person}{Takeo Igarashi}.}
  \bibinfo{year}{2021}\natexlab{}.
\newblock \showarticletitle{Interactive Exploration-Exploitation Balancing for
  Generative Melody Composition}. In \bibinfo{booktitle}{\emph{26th
  International Conference on Intelligent User Interfaces}}
  \emph{(\bibinfo{series}{IUI '21})}. \bibinfo{publisher}{Association for
  Computing Machinery}, \bibinfo{address}{New York, NY, USA},
  \bibinfo{pages}{43–47}.
\newblock
\showISBNx{9781450380171}
\urldef\tempurl%
\url{https://doi.org/10.1145/3397481.3450663}
\showDOI{\tempurl}


\end{thebibliography}
\end{document}